\documentclass{aa}

\usepackage{graphicx}
\usepackage{float}
\usepackage{txfonts}
\usepackage{multirow}

\usepackage{amsmath}	
\usepackage{xcolor}
\usepackage{url}
\usepackage{booktabs}
\usepackage{pifont}
\usepackage[normalem]{ulem}
\newcommand{\cmark}{\ding{51}}%
\newcommand{\xmark}{\ding{55}}%

\usepackage[draft=false]{hyperref}

\defcitealias{Anderson2022}{Anderson~et~al.~submitted}

\begin{document}

\title{Family dispute: do Type IIP supernova siblings agree on their distance?}

\author{ G.~Cs\"ornyei
		\inst{1,2}
		\and
		C.~Vogl
		\inst{1,3}
		\and
 		S.~Taubenberger
 		\inst{1,2}
  		\and
  		A.~Flörs
  		\inst{4}
    	\and
  		S.~Blondin
  		\inst{5}
   		\and
  		M.~G.~Cudmani
  		\inst{1,6,7}
  		\and
  		A.~Holas
  		\inst{8,9}
  		\and
  		S.~Kressierer
  		\inst{6,7}
   		\and
 		B.~Leibundgut
 		\inst{2,6}
 		\and
 		W.~Hillebrandt
  		\inst{1,2,3}
        }
 		 
\institute{	Max-Planck-Institute für Astrophysik, Karl-Schwarzschild-Str. 1, 85748 Garching, Germany			
		\and
			Physics Department, Technische Universit\"at M\"unchen, James-Franck-Str. 1, 85748 Garching, Germany
		\and
		    Exzellenzcluster ORIGINS, Boltzmannstr. 2, 85748 Garching, Germany
    	\and
		    GSI Helmholtzzentrum für Schwerionenforschung, Planckstraße 1, 64291 Darmstadt, Germany
		\and
		    Aix Marseille Univ, CNRS, CNES, LAM, Marseille, France
		\and
			European Southern Observatory, Karl-Schwarzschild-Str. 2, 85748 Garching, Germany
		\and
		    School of Natural Sciences, Technische Universität München, James-Franck-Str. 1., 85748 Garching, Germany
		\and
		    Zentrum f\"{u}r Astronomie der Universit\"{a}t Heidelberg, Institut für Theoretische Astrophysik, Philosophenweg 12, 69120 Heidelberg, Germany
		\and
		    Heidelberg Institute for Theoretical Studies, Schloss-Wolfsbrunnenweg 35,  69118 Heidelberg, Germany
}

\date{Received XXX}

\abstract
{Type II supernovae provide a direct way to estimate distances through the expanding photosphere method, which is independent of the cosmic distance ladder. A recently introduced Gaussian process-based method allows for a fast and precise modelling of spectral time series, which puts accurate and computationally cheap Type II-based absolute distance determinations within reach.}
{The goal of the paper is to assess the internal consistency of this new modelling technique coupled with the distance estimation empirically, using the spectral time series of supernova siblings, i.e. supernovae that exploded in the same host galaxy.}
{We use a recently developed spectral emulator code, which is trained on \textsc{Tardis} radiative transfer models and is capable of a fast maximum likelihood parameter estimation and spectral fitting. After calculating the relevant physical parameters of supernovae we apply the expanding photosphere method to estimate their distances. Finally, we test the consistency of the obtained values by applying the formalism of Bayes factors.}
{The distances to four different host galaxies were estimated based on two supernovae in each. The distance estimates are not only consistent within the errors for each of the supernova sibling pairs, but in the case of two hosts they are precise to better than 5\%. The analysis also showed that the main limiting factor of this estimation is the number and quality of spectra available for the individual objects, rather than the physical differences of the siblings.}
{Even though the literature data we used was not tailored for the requirements of our analysis, the agreement of the final estimates shows that the method is robust and is capable of inferring both precise and consistent distances. By using high-quality spectral time series, this method can provide precise distance estimates independent of the distance ladder, which are of high value for cosmology.}


\maketitle

\section{Introduction}

Sibling supernovae are transients that exploded in the same host galaxy. Given that these supernovae occurred at essentially the same distance from us, they allow us to test distance estimation methods and investigate their systematics empirically. Such consistency checks have been performed in the literature for pairs of Type Ia supernovae (see e.g. \citealt{Burns2020, Scolnic2020}) and for a pair of a Type Ia and a Type IIP supernova \citep{Graham2022} recently, yielding good matches. However, no similar tests were conducted on Type II supernovae specifically yet. 

Previously, Type II supernova siblings were analysed with different goals. For example, \cite{Poznanski2009} used such siblings to check the colour term in their standardizable candle model under the assumption that they share the same distance. More recently, \cite{Vinko2012} and \cite{Szalai2019} used a Type IIP-IIb and a IIP-IIP pair respectively to constrain the distances to their hosts better than with a single transient. In contrast, here we perform the first dedicated distance consistency check for Type IIP supernova siblings.


Type II supernovae (SNe II) correspond to the final gravitational collapse of massive ($\geq 8 M_{\odot}$), red or blue supergiant stars, which is supported by pre-explosion images \citep{Smartt2009} and theoretical models \citep{Tinyanont2021}. These supernovae are historically subdivided into two main classes, Type IIP (plateau) and Type IIL (linear), based on their light curves (e.g. \citealt{Patat1994}). However, there are several indications that these objects can be rather explained as a continuous distribution, as opposed to distinct classes \citep{Anderson2014,Sanders2015,Galbany2016,Morozova2017,Pessi2019}. 

Due to their high intrinsic luminosity and the fact that these supernovae are the most frequent stellar explosions in the Universe \citep{Li2011}, Type II SNe make for excellent distance indicators. To date, mainly two methods have been used to estimate the distances to Type II SNe: the expanding photosphere method (EPM, \citealt{Kirshner1974}), which is a geometric technique relating the photospheric radius of the SN to its angular size, and the standardized candle method (SCM, \citealt{Hamuy2002}), which is based on an empirical relation between the photospheric expansion velocity and the plateau luminosity of the supernovae. However, only the EPM provides a distance estimation that does not require any external calibration, as opposed to the SCM and other distance ladder formalism methods. It is independent of any other distance measurements and hence of the cosmic distance ladder. Owing to these advantages, it has already been applied to various sets of SNe to derive the Hubble constant (see e.g. \citealt{Schmidt1992}). The classical EPM analysis is, however, prone to several uncertainties as, for example, shown by \cite{Jones2009}: the results can be subject to systematic differences depending on which atmospheric model and thus dilution factors are assumed (e.g. \citealt{E96} or \citealt{DH05}) or which photometric passband set is used for the photometry. To avoid such issues, as pointed out by \cite{DH05}, one is required to carry out the EPM based on the radiative transfer-based modelling of the spectra and estimate the input parameters through these models. We call this augmented version the tailored-EPM analysis, which bears more similarities to the Spectral-fitting Expanding Atmosphere Method (SEAM) introduced by \cite{Baron2004}. This step not only allows for a more precise estimation of physical parameters but also avoids the detour of choosing the dilution factors for the EPM.

Although the number of available spectroscopic observations has grown significantly over the past years, the spectral modelling remains a time-consuming and laborious process. To change this situation, \cite{Vogl2020} developed an emulator based on spectra calculated with the \textsc{Tardis} radiative transfer code \citep{TARDIS}, which allows for a maximum likelihood parameter estimation and modelling of the spectral time series several orders of magnitude faster than the conventional methods. To showcase the code, \cite{Vogl2020} performed the tailored-EPM analysis for SNe 1999em and 2005cs, showing that a few per cent precision in the derived distance can be achieved. 

Here, we attempt to further test the method and investigate its internal consistency empirically by applying it to sibling supernovae. In their case, the maximum possible separation between the siblings is set by the line-of-sight extension of the thick disk of their host galaxies. For face-on galaxies (such as all the hosts in our sample), this is at most about 10 kpc \citep{Gilmore1983}. At a distance of $\sim$6 Mpc, for the closest host galaxy in our sample, this corresponds only to a maximum relative error of 0.2\%. Hence, the EPM should yield the same distance for the sibling pairs within the uncertainties. As a result, siblings provide a simple empirical way to assess the consistency and robustness of the algorithm: they allow us to test whether we find the same distances for the pairs even though the underlying conditions (such as the metallicity, the mass of the progenitor, the amount of ejecta-CSM interaction, the reddening), as well as the overall data quality and the level of calibration are different, and whether the inferred uncertainties are reasonable.

The paper is structured as follows. In Sect.~\ref{sec:Data} we give a brief overview of the data collected for this study. Sect.~\ref{sec:Methods} describes the calibration steps we applied to the data to achieve a similar level of calibration for all objects, provides the background of the emulator-based modelling, and finally outline the distance estimation. Sect.~\ref{sec:Results} shows the results of the modelling for the individual host galaxies, while Sect.~\ref{sec:Discuss} discusses these results and details the consistency check of the method. In Sect.~\ref{sec:Summary} we summarize and conclude the paper.


\section{Data}
\label{sec:Data}
To obtain a set of objects compatible with the goals of this study, we filtered the catalogue of known supernovae through the Open Supernova Catalog\footnote{\url{https://sne.space/}} (OSC, \citealt{OSC}). Apart from looking for supernovae that exploded in the same host, we posed further constraints on the individual objects to make sure that our method could be applied to them: the objects should have at least one observation in their early photospheric phase (more precisely, in the epoch range of 10 to 35 days with respect to the time of explosion, see Sect. \ref{sec:SpecMod}) to be compatible with our radiative-transfer modelling, a well-covered light curve during these epochs (optimally, in multiple bands for calibration purposes, see Sect. \ref{sec:Calib}) and a well-constrained time of explosion from either non-detections or from fitting of the rise of the light curve.

By filtering the catalogue, we found that SN~IIP pairs in four host galaxies met all the conditions that we described above: NGC~772, NGC~922, NGC~4303 (M~61), and NGC~6946 (Fig.~\ref{fig:hosts}). To retrieve the data for the supernovae in these hosts, we made use of the OSC and WISeREP\footnote{\url{https://www.wiserep.org/}} \citep{wiserep}. The properties of the final dataset are summarized in Table~\ref{tab:SNProps}.

\begin{table*}
\begin{center}
    \begin{tabular}{c c c | c c c c}
Host & $z_{\textrm{helio}}$ & $E(B-V)_{\textrm{MW}}$ & SN & $N_{\textrm{spec}}$ & $N_{\textrm{phot}}$ (bands) & References\\
\hline
\multirow{2}{*}{M61} & \multirow{2}{*}{0.00522} & \multirow{2}{*}{0.0193} 
                            & 2008in & 6 & 22 (BVRI) & \cite{2008in, deJaeger2019} \\
                    &    &  & 2020jfo & 7 & 21 ($gri$) & \cite{2020jfo, Sollerman2021} \\
\hline
\multirow{2}{*}{NGC 772} & \multirow{2}{*}{0.00825} & \multirow{2}{*}{0.0623} 
                            & 2003hl & 1 & 36 (BVRI) & \cite{2003hl, Faran2014} \\
                    &    &  & 2003iq & 4 & 44 (BVRI) & \cite{2003iq, Faran2014} \\
\hline
\multirow{2}{*}{NGC 922} & \multirow{2}{*}{0.01028} & \multirow{2}{*}{0.0163} 
                            & 2002gw & 3 & 34 (BVI) & \cite{2002gw, Galbany2016} \\
                    &    &  & 2008ho & 2 & 11 (BV) & \cite{2008ho}; \citetalias{Anderson2022} \\
\hline
\multirow{2}{*}{NGC 6946} & \multirow{2}{*}{0.00013} & \multirow{2}{*}{0.2967} 
                            & 2004et & 6 & 40 (BVRI) & \cite{2004et, Faran2014} \\
                    &    &  & 2017eaw & 8 & 96 (BVRI) & \cite{2017eaw, Szalai2019} \\
\hline
    \end{tabular}
\end{center}
\caption{Summary table of the compiled SN sample, along with important properties of the host, namely its heliocentric redshift (as adopted from the OSC) and reddening caused by the dust in the Milky Way \citep{Schlafly2011}. References are provided for the discovery and photometry of the objects; for details on the spectroscopy, see Sect.~\ref{sec:Results}.}
\label{tab:SNProps}
\end{table*}


\section{Methods}
\label{sec:Methods}
\subsection{EPM and spectral modelling inputs}
\label{sec:Calib}

Before we could fit the spectral time series and perform the EPM analysis, we needed to obtain the necessary input data: an estimate of the time of explosion $t_0$, photometry interpolated to the spectral epochs, and flux-calibrated spectra. We describe how we inferred $t_0$ from a parametric fit in Sect.~\ref{sec:time_explosion}. Section~\ref{sec:lc_interpolation} explains the Gaussian process (GP) interpolation of the light curves and Sect.~\ref{sec:flux_calibration} details how we used the interpolated magnitudes to recalibrate the spectra.

\subsubsection{Time of explosion}
\label{sec:time_explosion}

The time of explosion is a crucial parameter in the EPM. It sets the size of the photosphere and thus the model luminosity. Any error in $t_0$ causes an error in the distance. In case the distance to a supernova is estimated using a single epoch, the uncertainty in the time of explosion $t_0$ translates directly into the distance uncertainty:

\begin{equation}
\frac{\Delta D}{D} \approx \sqrt{\left(\frac{\Delta t_0}{t - t_0}\right)^2 + \left(\frac{\Delta (\Theta/v)}{\Theta/v}\right)^2}
\end{equation}

based on linear relative error propagation, in accordance with the equations shown in Sec.~\ref{sec:EPM}. The parameters with the $\Delta$ denote the uncertainty of the given values. Assuming a 10\% error on the $\Theta/v$ measurement for a single spectrum at the epoch of 20 days, a $t_0$ uncertainty of $\pm 2$ days would yield $\sim 14$\% error on the final distance, which is too large for our purposes.

Analyzing multiple spectral epochs helps limiting the final uncertainties of the fit parameters, partially owing to the additional constraining effect exerted on $t_0$ by the EPM regression. For example, having four high-quality observations in the epoch range of 10 to 25 days along with the above $t_0$ uncertainty of 2 days would yield an EPM distance error of about $\sim 11$\%. While the improvement in precision is significant compared to the above case, bringing it to the required levels would require a factor of a few more spectral epochs, which is rarely available. This demonstrates that even with a sufficiently large number of observations having a well-constrained prior estimate on $t_0$ is crucial for the distance determination.



We can use the constraints on $t_0$ from the early light curve to significantly reduce these uncertainties. Also, if the time of explosion is known to a high precision (to an uncertainty of less than a day ideally), independently of the EPM analysis, even a single spectral epoch is enough to obtain a meaningful distance. Observationally, the common approach for estimating the time of explosion has been to take the midpoint between the first detection and last non-detection (e.g. \citealt{Gutierrez2017}). This is not accurate enough for our analysis. The method does not take into account the depth of the non-detections in comparison to the first detection; this can bias the estimated time of explosion and thus the distance. Also, the approach neglects the information from data on the rise of the light curve.

To minimize these possible biases and improve the precision, we estimate the time of explosion for each SN through the fitting of their early light curves (the initial plateau, i.e. the first $10-40$~days, depending on the individual SNe). Following \cite{Ofek2014} and \cite{Rubin2016}, we fit the flux $f$ in band $W$ with a model 

\begin{equation}
    f_W(t) = f_{m, W} \left[1 - \exp{\left(-\frac{t-t_0}{t_{e, W}}\right)}\right],
\end{equation}

where $t$ is the time, $t_0$ is the time of explosion, $f_{m, w}$ is the peak flux, and $t_{e,W}$ is the characteristic rise time in the particular band. Whenever the data allowed, we carried out this fitting for multiple photometric bands simultaneously to increase the accuracy of the method. In this joint fit, each of the different bands had its own $f_{m,W}$ and $t_{e,W}$, but the fits were connected through the global $t_0$ value, which was the same for all bands. To improve the fitting, we introduced two additional constraints: (i) we took the depth of the non-detections into account by placing the constraint that the model flux should not exceed the limiting flux (but the light curves were allowed to extend to times before the non-detection), and (ii) we required that the characteristic time scale $t_e$ for the light curve rise increases with wavelength for a given supernova, as it was found previously \citep[see, e.g.,][]{Gonzalez2015}. This way, even the $t_e$ values, which otherwise correspond to their own band, were constrained globally in the joint fit.

To fit the model to the set of light curves in different bands simultaneously, we applied the \texttt{UltraNest}\footnote{\url{https://johannesbuchner.github.io/UltraNest/}} package \citep{Buchner2021}, which allows for Bayesian inference on complex, arbitrarily defined likelihoods and which derives the necessary posterior probability distributions based on the nested sampling Monte Carlo algorithm MLFriends \citep{Buchner2016, Buchner2019}. We assumed a flat prior for each of the input parameters and used a Gaussian likelihood for the parameter inference. To account for a possible underestimation of the photometric errors, which can influence the inferred parameters, we extended the likelihood with an additional term corresponding to the error inflation as described in \cite{Hogg2010}. Using this procedure, we obtained a $t_0$ posterior distribution for each supernova. We then used the maximum of this distribution as our single "best estimate" value for $t_0$ to set the phase of the spectral observations for the fitting. The full distribution is used as a prior for the EPM analysis, as described in Sect.~\ref{sec:EPM}. This unique approach for obtaining $t_0$ in parallel to the EPM significantly enhances the precision of the distance measurements, as will be shown in Sect.~\ref{sec:Results}.

\subsubsection{Interpolated photometry}
\label{sec:lc_interpolation}

The epochs of spectral observations are not necessarily covered by individual photometric datapoints. To obtain photometry at all phases, we fit the light curves using GPs by applying the \texttt{george}\footnote{\url{https://george.readthedocs.io/en/latest/}} python package \citep{george}. GPs present a useful way to interpolate the light curves as they provide a non-parametric way of fitting, while taking into account the uncertainties of the datapoints. Following several works that already used GPs for the modelling of supernova light curves \citep{Inserra2018,Yao2020, Kangas2022}, we adopted covariance functions from the Mat\'{e}rn family for our calculations \citep{Rasmussen2006}. We chose a smoothness parameter of $3/2$ for our analysis. We attempted to keep the length scale of the best-fit GP curve high enough to retain its robustness against the scatter present in the datapoints. To achieve this, we omitted the rise of the light curve and the drop from the plateau from the fits. Figure \ref{fig:lcs} shows the interpolated light curves and Table~\ref{tab:magnitudes} lists the magnitudes at the spectral epochs. The interpolated magnitudes are direct inputs for the EPM, as described in Sect.~\ref{sec:EPM}.

\begin{figure*}
\centering
\includegraphics[width=0.495\linewidth]{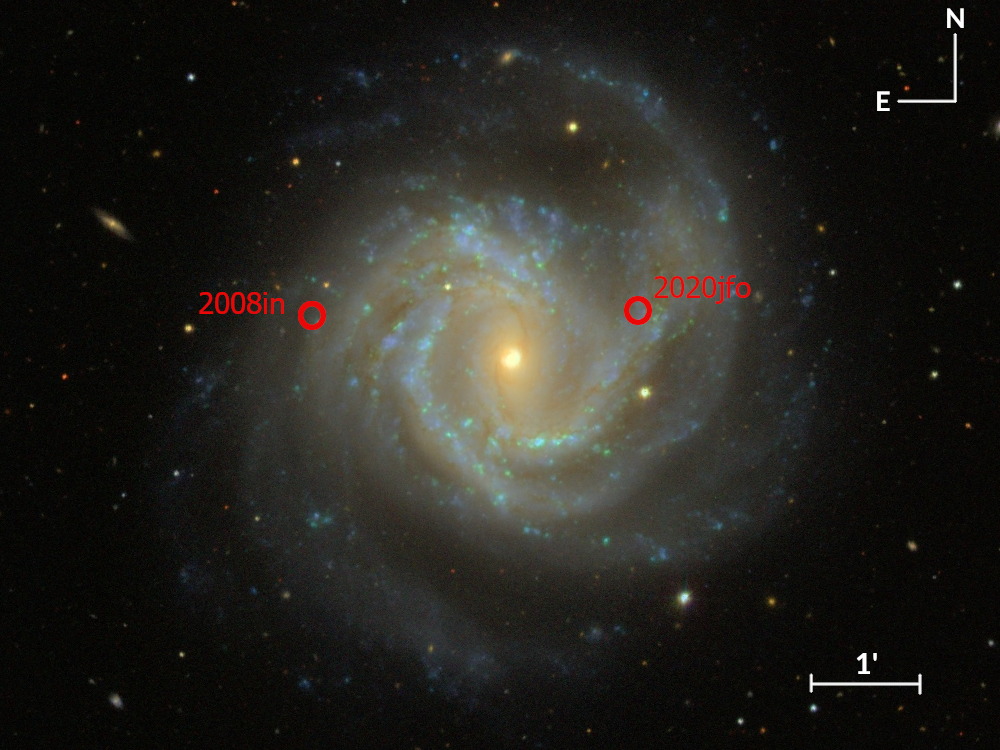}
\includegraphics[width=0.495\linewidth]{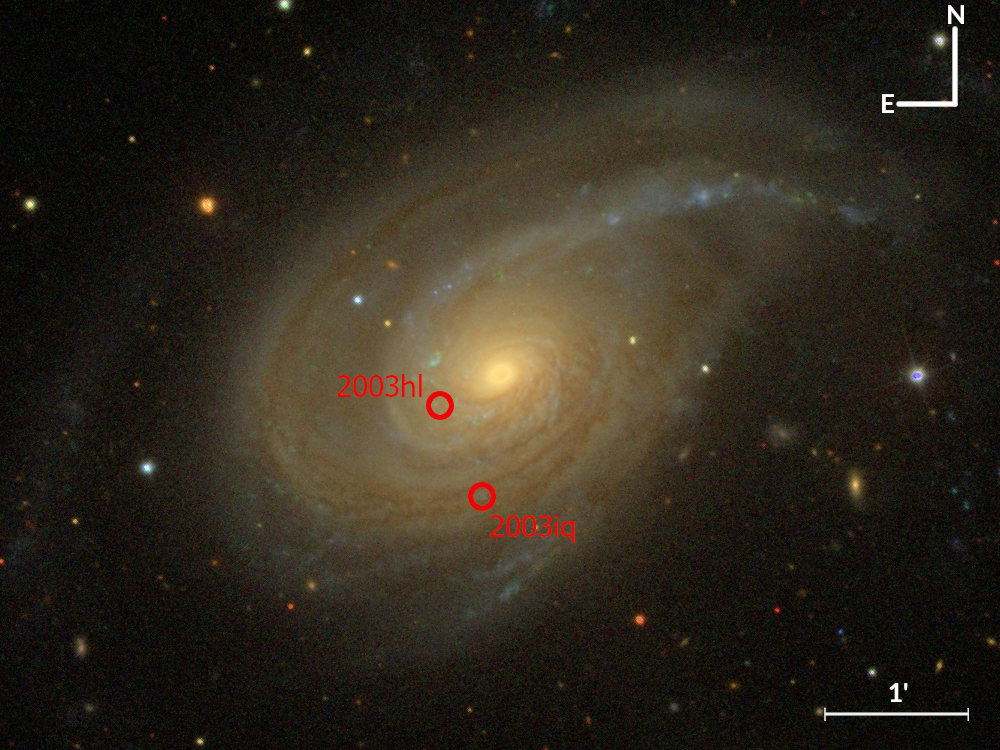}
\includegraphics[width=0.495\linewidth]{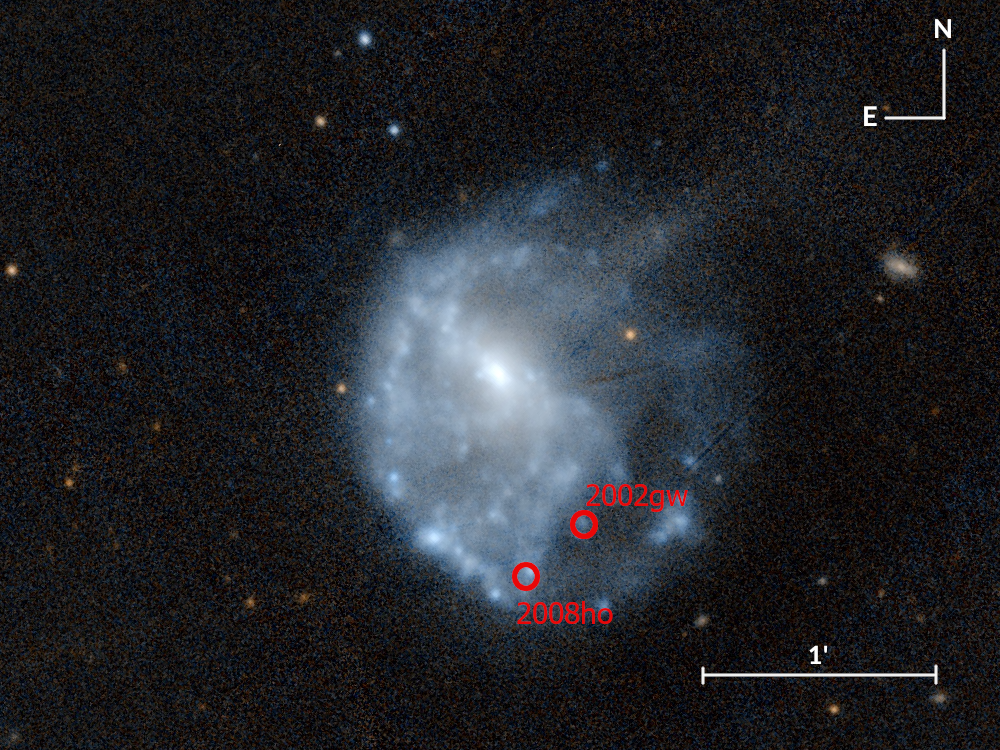}
\includegraphics[width=0.495\linewidth]{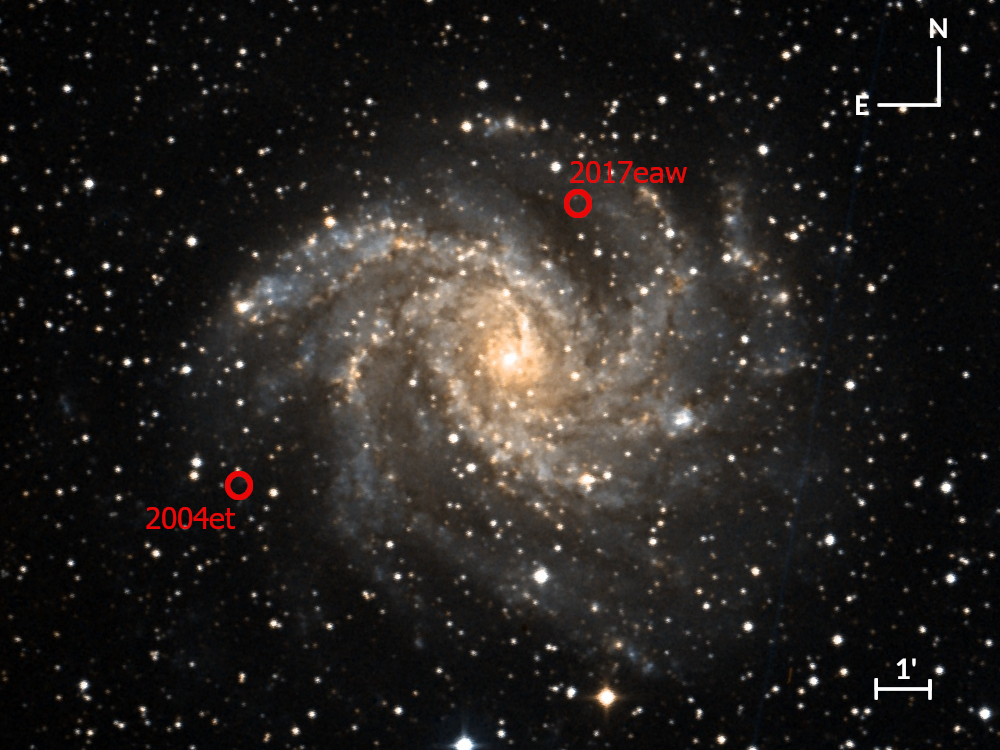}
\caption{Images of the sample host galaxies. \textbf{Top left:} SDSS DR9\protect \footnotemark image of M61, along with the positions of SN 2008in and 2020jfo. \textbf{Top right:} SDSS DR9 image of NGC 772, along with the positions of 2003iq and 2003hl. \textbf{Bottom left:} PanSTARRS DR1\protect \footnotemark image of NGC 922, along with the positions of 2002gw and 2008ho. \textbf{Bottom right:} DSS\protect\footnotemark image of NGC 6946, along with the positions of 2004et and 2017eaw.}
\label{fig:hosts}
\end{figure*}

\subsubsection{Flux calibrated spectra}
\label{sec:flux_calibration}

A reliable flux calibration is crucial for the spectral modelling and the determination of the host extinction. We therefore apply a linear flux correction to avoid possible biases in the analysis. This correction was based on the photometry: for every epoch, we calculated a set of synthetic magnitudes using the transmission curves from \cite{Bessell2012} (for $BVRI$ magnitudes) or \cite{ZTFfilts} (in case of ZTF photometry), and then compared these values to the corresponding interpolated magnitudes. The correction curve was then calculated by fitting the linear trend present in the ratios of synthetic and interpolated fluxes against the effective wavelengths of the passbands. Finally, the spectra were multiplied with the obtained correction trends. 
In some cases, additional more complex flux calibration steps were required, which are described for the individual supernovae separately.\addtocounter{footnote}{-3}
\stepcounter{footnote}\footnotetext{\url{https://www.sdss3.org/index.php}}
\stepcounter{footnote}\footnotetext{\url{https://archive.eso.org/dss/dss}}
\stepcounter{footnote}\footnotetext{\url{https://outerspace.stsci.edu/display/PANSTARRS/}}

\subsection{Spectral modelling}
\label{sec:SpecMod}
To fit the individual spectra, we applied the methodology introduced by \cite{Vogl2020}. The fitting method is based on a spectral emulator that predicts synthetic spectra for a set of supernova parameters. The training set of synthetic spectra was calculated with a modified version of the Monte Carlo radiative transfer code \textsc{Tardis} \citep{TARDIS, Vogl2019}, and is described in detail in \cite{Vogl2020}. To allow for a fast and reliable interpolation of model spectra for a given set of physical parameters, along with correct absolute magnitudes, an emulator was built on this training set with the following procedure: first, the dimensionality of the spectra was greatly reduced by the use of principal component analysis (PCA), then GPs were trained on the physical input parameters (photospheric temperature and velocity, $T_{\textrm{ph}}$ and $v_{\textrm{ph}}$, metallicity $Z$, time since explosion $t_{\textrm{exp}}$, and the exponent of the density profile $n = - \textrm{d}\ln \rho / \textrm{d} \ln r$) to interpolate spectra and to predict absolute magnitudes. As described in \cite{Vogl2020}, the emulator predicts spectra that in almost all cases match the simulations with a precision of better than 99\%, as measured by the mean fractional error. The original version of the emulator covered the early photospheric phase, from 6.5 to 22.5 days after explosion \citep{Vogl2020}. This was extended towards later epochs up until 38 days, and additional high-temperature models with an NLTE (non-local thermal equilibrium) treatment of He were also added \citep{Vogl2020b, Vasylyev2022, Vogl2022}. The 38-day upper end is set by the modelling limitations, since characteristically after this epoch time-dependent effects in the excitation and ionization-balance become important.\footnote{The choice of 38 days is arbitrary to a certain degree and arguments can be made for lower and higher values for the cutoff: time-dependent effects already play a minor role at earlier epochs, while single snapshot models are probably still accurate enough for slightly later epochs. Modelling such late epochs is more complex and time intensive, henceforth the emulator is limited to earlier times, where models can be calculated in single snapshot simulations.} The physical range covered by the emulator is summarized in Table~\ref{tab:ext_emulator}.

\begin{table}[t]
  \centering
  \resizebox{\columnwidth}{!}{
  \begin{tabular}{cccccccc}
   \toprule \toprule
     & $v_\mathrm{ph}$\,[km$\,\mathrm{s}^{-1}$]  & $T_\mathrm{ph}\,$[K] & $Z\,[Z_\odot]$ & $t_\mathrm{exp}\,$[days] & n & \multicolumn{2}{c}{NLTE} \\
    \midrule
    \multicolumn{6}{c}{$\mathbf{t_\mathrm{exp}\,}$ \textbf{< 16 days}} & \textbf{\small{H}} & \textbf{\small{He}} \\
    \midrule
    Min & 4500 & 7200  & 0.1 & 6.0 & 9 & \multirow{2}{*}{\cmark} & \multirow{2}{*}{\cmark}  \\
    Max & 12000 & 16000 & 3.0 & 16.0 & 26 \\
    \midrule
    \midrule
    \multicolumn{6}{c}{$\mathbf{t_\mathrm{exp}\,}$ \textbf{> 16 days}} & \textbf{\small{H}} & \textbf{\small{He}} \\
    \midrule
    Min & 3600 &  5800 & 0.1 & 6.5 & 6 & \multirow{2}{*}{\cmark} & \multirow{2}{*}{\xmark} \\
    Max & 10700 & 10000 & 3.0 & 40.0 & 16 \\
    \bottomrule
  \end{tabular}
  }
  \caption[Parameter range covered by the extended spectral emulator]{Parameter range covered by the extended spectral emulator. Depending on the spectral epoch, one of these two simulation sets was chosen for building the emulator.}
  \label{tab:ext_emulator}
\end{table}

To infer physical parameters, we used a Gaussian likelihood:

\begin{equation}
    \ln p(f_{\lambda}^{\textrm{obs}}|\theta_{\textrm{SN}} , E(B-V)) = - \frac{1}{2}(R^{\textrm{T}}C^{-1}R + \ln \textrm{det } C + N \ln 2\pi),
\end{equation}
where,
\begin{equation}
    R = f_{\lambda}^{\textrm{obs}} - f_{\lambda}(\theta_{\textrm{SN}}, E(B-V)),
\end{equation}

where $\theta_{\textrm{SN}} = (v_{\textrm{ph}}, T_{\textrm{ph}}, Z, t_{\textrm{exp}}, n)$ denotes the set of physical parameters, $f_{\lambda}^{\textrm{obs}}$ and $f_{\lambda}$ denote the observed and the reddened emulated spectra, respectively, $N$ is the number of spectral bins and $C$ is the bin-to-bin covariance matrix (see, e.g., \citealt{Czekala2015}). The matrix $C$ is important for the inference. It should account for uncertainties in the data and the model, and capture the correlations of these uncertainties across wavelength; otherwise, we would significantly underestimate the parameter uncertainties \citep[see,][]{Czekala2015}. Constructing a matrix with these properties is a challenging, unsolved problem.

We thus used a simple diagonal matrix with constant values as in \citet{Vogl2020}. Since we cannot infer reasonable uncertainties under these circumstances, we performed only maximum likelihood estimation for the parameters. Throughout the fitting, the $t$ epoch of each spectrum was fixed. We treated the reddening towards the supernova separately from the other physical parameters: instead of directly including it in the likelihood function, we set up a grid of reddened spectra corresponding to various $E(B-V)$ values, and then evaluated the likelihood for each separately. Apart from reducing computational time, this allowed us to quantify the uncertainties caused by the reddening using the treatment described below. We applied the reddening correction according to the \cite{CCM1989} law with $R_V = 3.1$. For the lower limit of the $E(B-V)$ grid, we always assumed the Galactic colour excess towards the supernova, which was determined based on the \cite{Schlafly2011} dust maps. The best-fit $E(B-V)$ was chosen as the average of the $E(B-V)$ values that resulted in the lowest $\chi^2$ for the individual spectra.

Apart from calculating the best-fit physical parameters, we also evaluated the angular diameter values for every $E(B-V)$ gridpoint, to allow us to quantify the distance uncertainties resulting from the unknown amount of host reddening as described in the next section. 

\begin{figure}
    \centering
    \includegraphics[width = \linewidth]{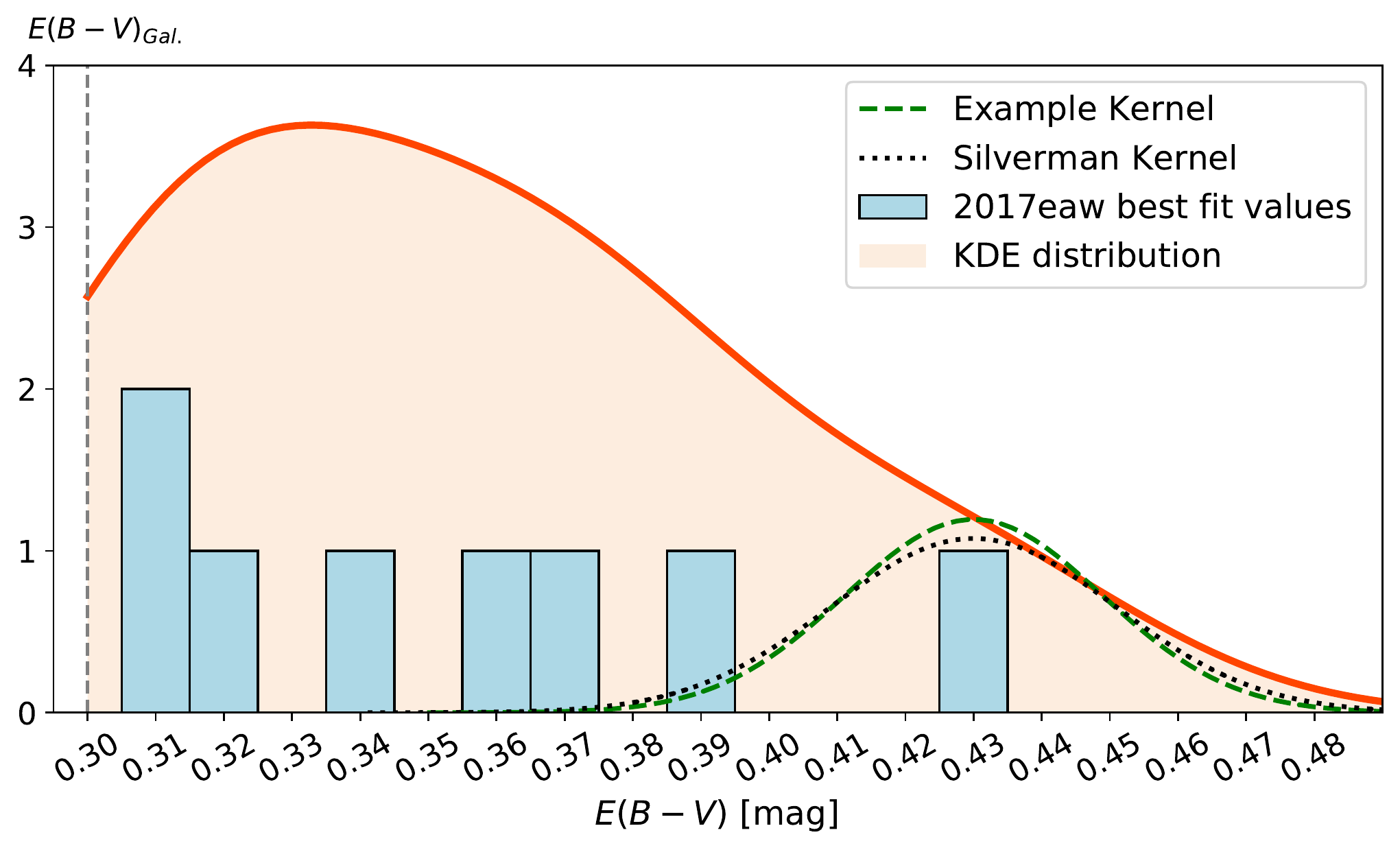}
    \caption{The distribution of best-fitting reddening estimates for all the epochs of SN~2017eaw (blue bars), and the constructed KDE distribution (background shape). The dashed grey line at $E(B-V) = 0.3$ mag denotes the Galactic reddening towards NGC~6946 based on the \cite{Schlafly2011} dust map, which sets the lower limit for the KDE. The green and grey curves show the Gaussian kernels for a single observation obtained by our setting and the Silverman's Rule respectively. A normalization was applied on the KDE histogram for a better comparison. For more details on this supernova, see Sect.~\ref{sec:6946}.}
    \label{fig:kde}
\end{figure}

\subsection{Distance determination}
\label{sec:EPM}
To infer the distances $D$ for the supernovae, we used a variant of the tailored-EPM method \citep{DH06, Dessart08, Vogl2020}. As a first step, the photospheric angular diameter of the supernova ($\Theta = R_{\textrm{ph}}/ D$, where $R_{\textrm{ph}}$ denotes the radius of the photosphere) had to be inferred for each of the spectral epochs. The predicted apparent magnitude for a passband $S$ depends on $\Theta$ as follows:

\begin{equation}
\begin{split}
    &m_S = M_S^{\textrm{ph}}(\Sigma^{*}) - 5 \log (\Theta) + A_S\\
    &M_S^{\textrm{ph}} = M_S + 5 \log \frac{R_{\textrm{ph}}}{10 \textrm{ pc}}
\end{split}
\end{equation}
Here, $\Sigma^{*}$ denotes the set of physical parameters corresponding to the best fit, $M_S^{\textrm{ph}}$ is the absolute magnitude predicted by the radiative transfer model at the position of the photosphere and $A_S$ denotes the broadband dust extinction in the bandpass. $M_S$ denote the absolute magnitude as defined by the distance modulus formula. The absolute magnitudes are transformed with above formula so that the variations in the size will have no effect. With this definition, we determined the best-fitting angular diameter $\Theta^{*}$ by minimizing the square difference between the observed $m_S^{\textrm{obs}}$ and model magnitudes:

\begin{equation}
    \Theta^{*} = \arg \min_{\hspace*{-15pt}\Theta} \sum_{\textrm{S}}\left(m_{\textrm{S}} - m_{\textrm{S}}^{\textrm{obs}}\right)^2
\end{equation}

\begin{figure*}
    \centering
    \includegraphics[width=0.500\linewidth]{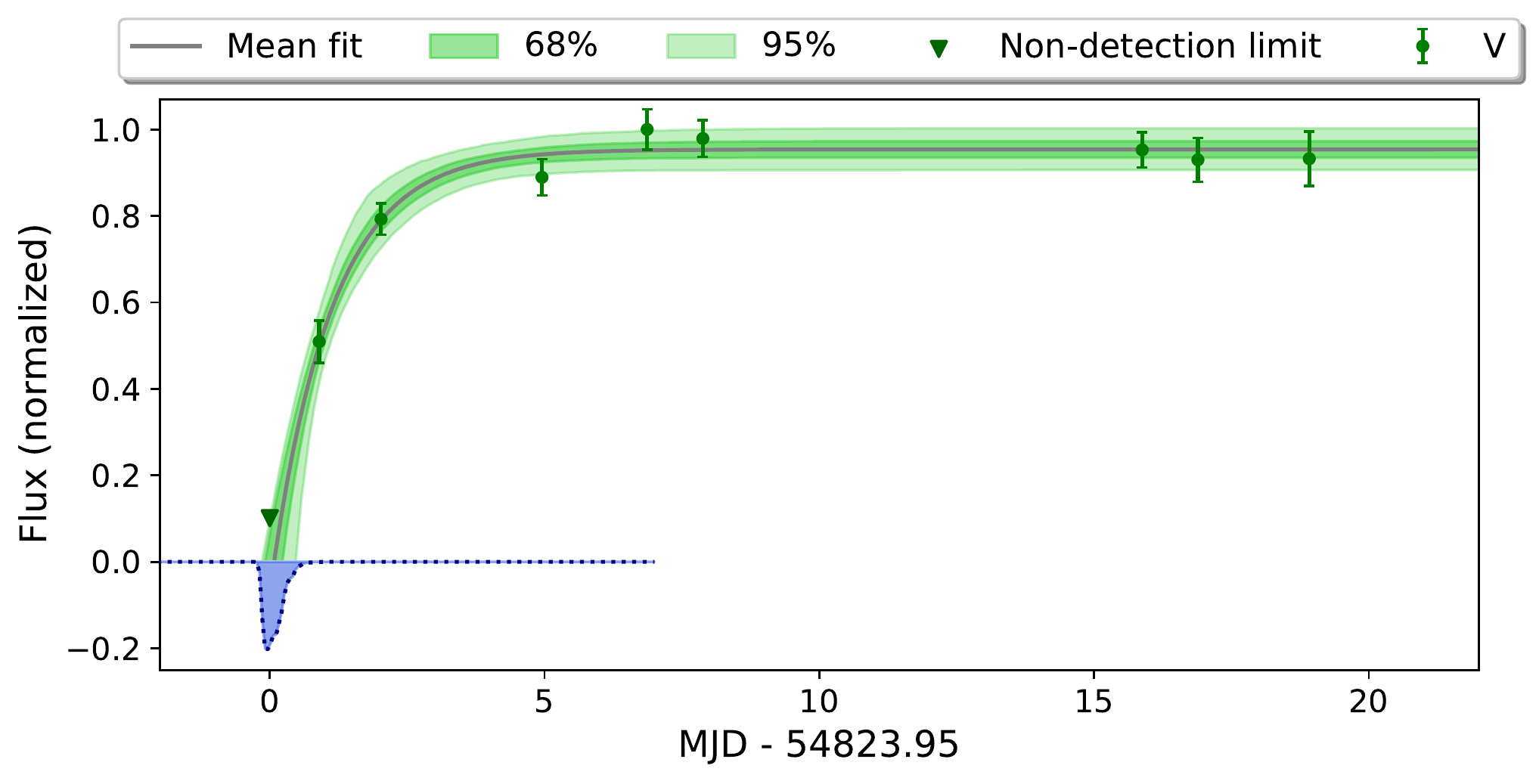}
    \includegraphics[width=0.492\linewidth]{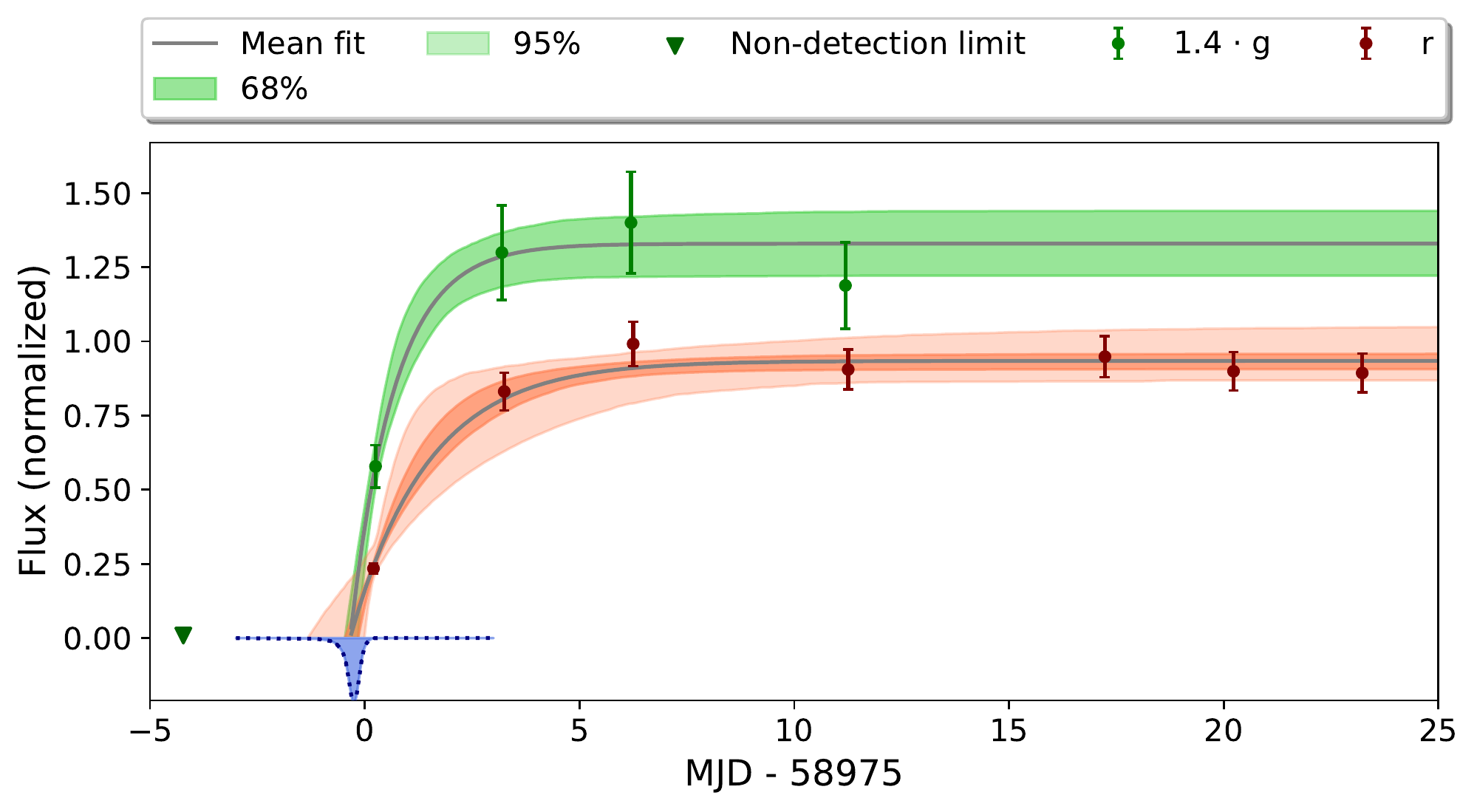}
    \caption{Exponential fit on the ROTSE light curve of SN 2008in (\textbf{left}), and the ZTF early light curves of SN 2020jfo (\textbf{right}). The blue shaded areas shows the obtained $t_0$ posteriors. The black curves show the mean fit, i.e. the fit that results by taking the mean of the posterior distributions. In case of SN 2020jfo, the $g$-band light curve was further rescaled after the normalization to improve clarity on the figure. The shaded regions denote the 68$\%$ and 95$\%$ confidence intervals.}
    \label{fig:2020jfo-2008in_lc}
\end{figure*}

\begin{figure*}
    \centering
    \includegraphics[width=\linewidth]{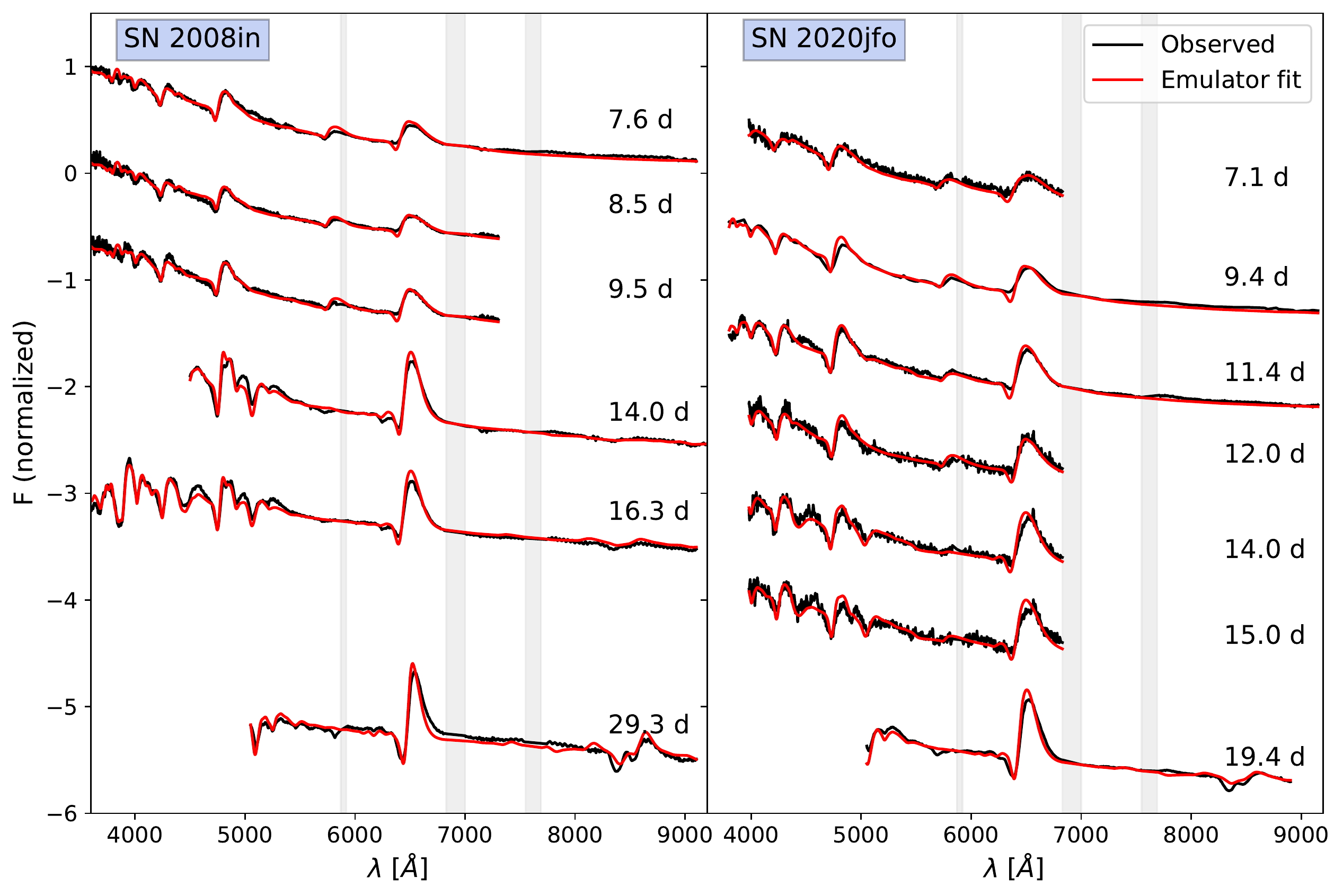}
    \caption{Spectral time series of SNe 2008in \citep{Roy2011,Hicken2017} and 2020jfo \citep{Sollerman2021}, along with our fits. The grey shaded areas denote telluric regions.}
    \label{fig:NGC4303_fits}
\end{figure*}

\begin{figure*}
    \centering
    \includegraphics[width=\linewidth]{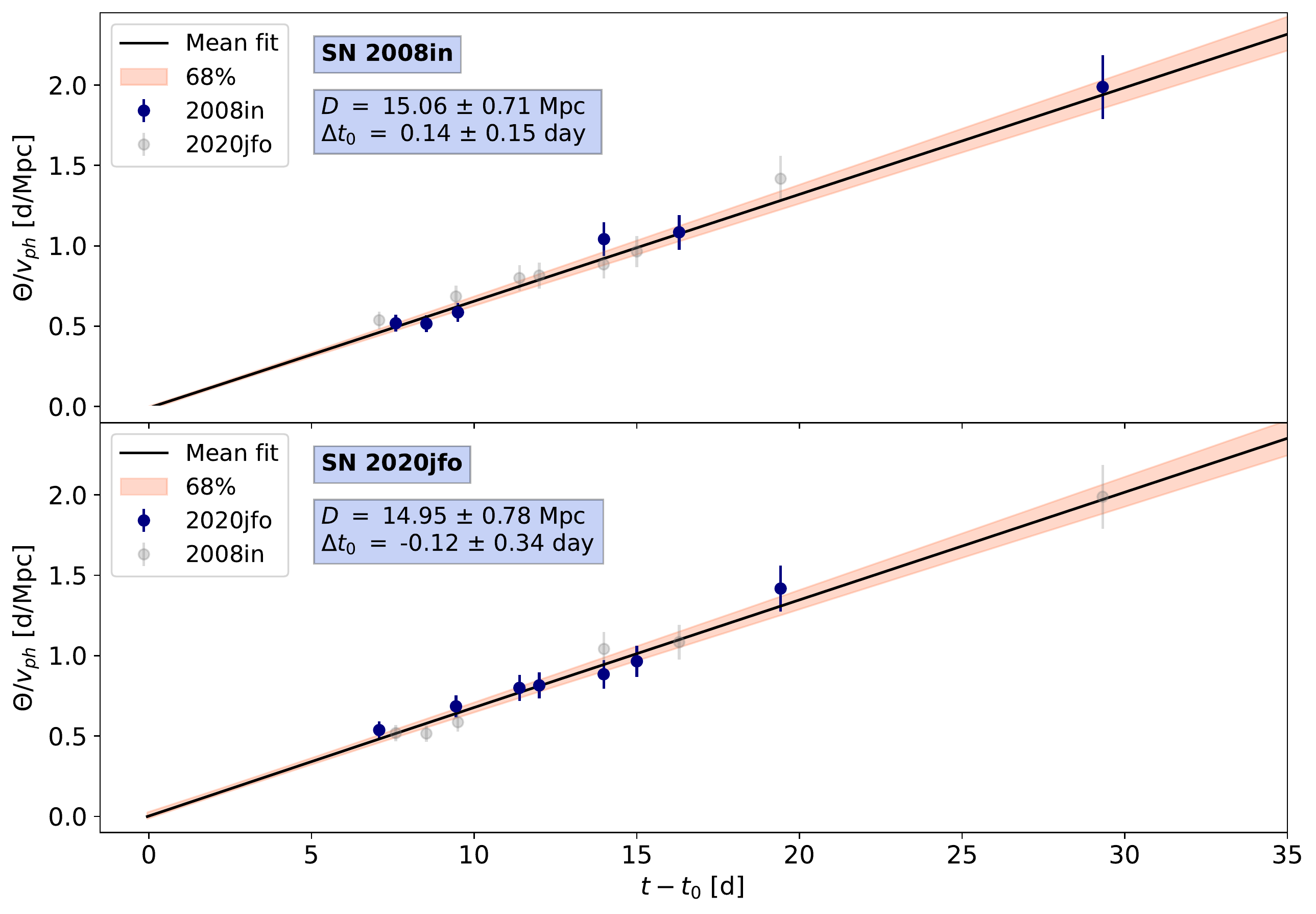}
    \caption{Evolution of $\Theta / v_{\textrm{ph}}$ for SNe 2008in (\textbf{top} panel) and 2020jfo (\textbf{bottom} panel). The derived distance $D$ and the final time of explosion estimate (with respect to the initial light curve fit based
    value) $\Delta t_0$ are displayed in the top left corner of each panel. The shaded region shows the 68$\%$ confidence interval of the fit. The displayed uncertainties denote the $1\sigma$ errors on $\Theta / v_{\textrm{ph}}$.}
    \label{fig:NGC4303_EPM}
\end{figure*}

Finally, assuming homologous expansion $R_{\textrm{ph}} = v_{\textrm{ph}}t$ (which is well motivated by observations and models for normal SNe IIP, see e.g. \citealt{Woosley1988, Dessart2005}) we determined the distance to the supernova and its time of explosion through a Bayesian linear fit to the ratios of the angular diameters and the photospheric velocities ($\Theta / v_{\textrm{ph}}$) against time $t$. In the fit, we assumed Gaussian uncertainties for $\Theta / v_{\textrm{ph}}$ of 10\% of the measured values for a given colour excess as in \citet{DH06,Dessart08,Vogl2020}. We used a flat prior for the distance, whereas for the time of explosion we used a normalized histogram of the $t_0$ posterior from the early light curve fit as the prior. However, instead of applying the standard $\chi^2$-based likelihood, we used a different approach to account for the correlated errors caused by the reddening.

This is important since the reddening can affect the final EPM distance, as $E(B-V)$ influences the $\Theta$ measurements through the de-reddening of the observed magnitudes and changes in the best-fit parameters such as the photospheric temperature. Hence, the uncertainty of the reddening introduces a correlated error to the final $\Theta / v_{\textrm{ph}}$ measurements, which transitions over to the distance obtained from the EPM fit. Depending on the exact model, higher extinction usually leads to shorter inferred distances. To take this into account, we extended the likelihood with the uncertainty of the reddening by applying kernel density estimation.

Kernel density estimation (KDE) is an effective tool for approximating the underlying probability density distribution of an observable using only a number of realizations. We use the \texttt{scipy} KDE implementation assuming Gaussian kernels \citep{scipy} to estimate an underlying distribution for the reddening values obtained from the individual spectral fits. The distributions calculated by KDE depend on the bandwidth of the individual Gaussian kernels. We set the bandwidth to a constant value of 0.025. This choice was based on empirical comparison with Silverman's Rule \citep{Silverman1986}, which is regularly used for KDE: for multiple epochs, the bandwidth estimate obtained using Silverman's Rule is in agreement with our preset value, hence the resulting distributions match in the two cases. However, Silverman's Rule cannot be used for a single epoch, or when a sole value of $E(B-V)$ is favoured by all fits; in these cases, our bandwidth choice still provided a realistic KDE. Finally, we set the lower limit of the estimated distribution to the Galactic reddening based on the \cite{Schlafly2011} map, to exclude non-physical cases. One of the obtained KDE distributions can be seen in Fig.~\ref{fig:kde}.

To incorporate the correlated uncertainty in the fit, we first drew a large number of reddening samples using the obtained KDE. Then, a respective sample of $\Theta / v_{\textrm{ph}}$ values was generated based on the $\Theta / v_{\textrm{ph}} - E(B-V)$ linear interpolation, and by adding a random offset to each value assuming 10\% uncertainty. This sample not only contained the $\Theta / v_{\textrm{ph}}$ values for the relevant reddening values only, but it also carried information about the correlated errors present in them. To represent these distributions and set up the final EPM regression, we applied a multi-dimensional Gaussian KDE on the set of $\Theta / v_{\textrm{ph}}$ values, and then used the corresponding probability density function as the likelihood for the fitting. We then evaluated this likelihood using \texttt{UltraNest}: at each step, the sampler drew an assumed distance and time of explosion value, for which the model $(\Theta / v_{\textrm{ph}})^{*}$ were calculated, and then the log probability density function was evaluated. As a result, we obtained a posterior distribution for the distance that includes the correlated error introduced by the uncertainty in the reddening. 

\section{Results}
\label{sec:Results}

In this section, we present the distances obtained for the individual host galaxies based on their two sibling supernovae. 

\begin{figure*}
    \centering
    \includegraphics[width=0.490\linewidth]{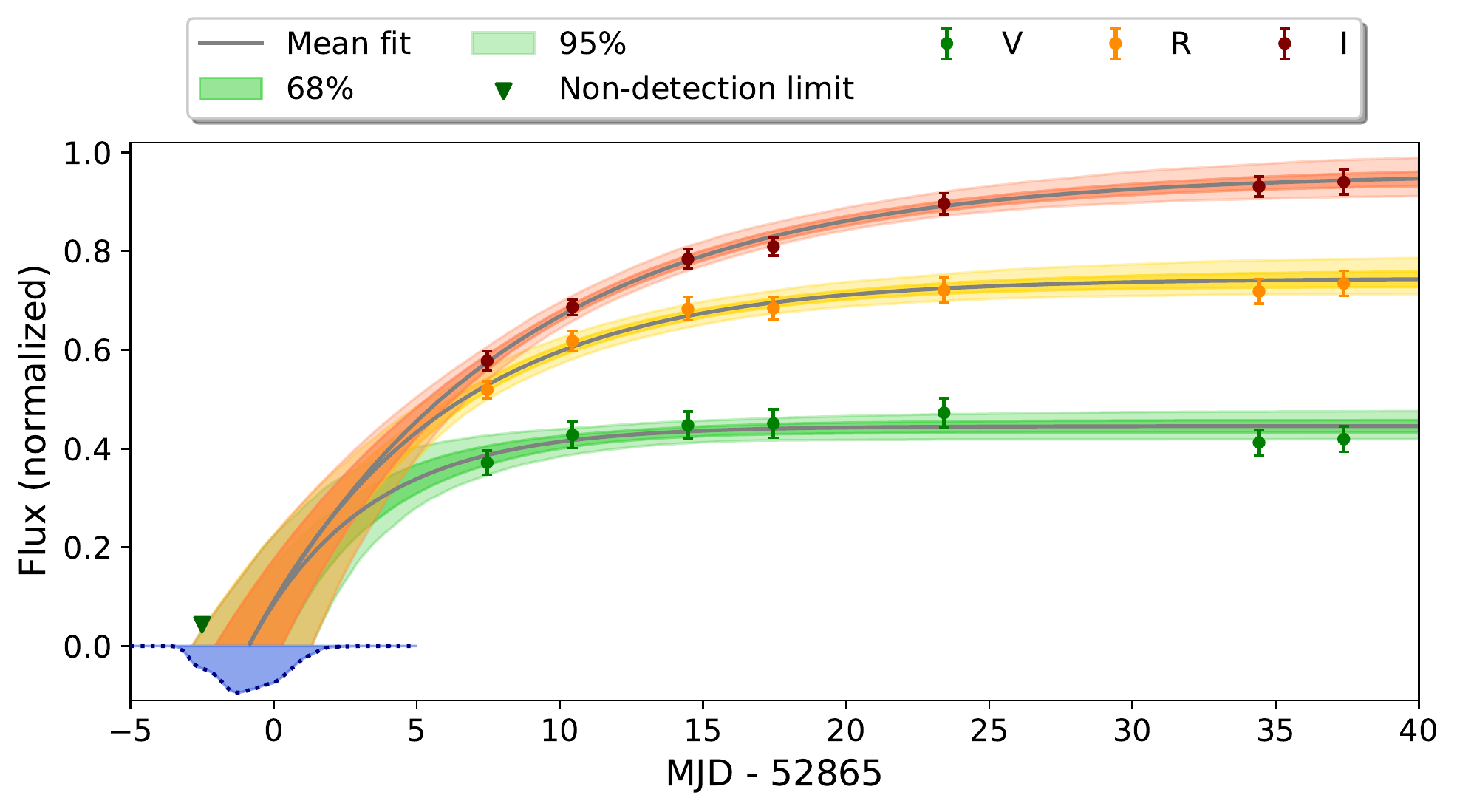}
    \includegraphics[width=0.500\linewidth]{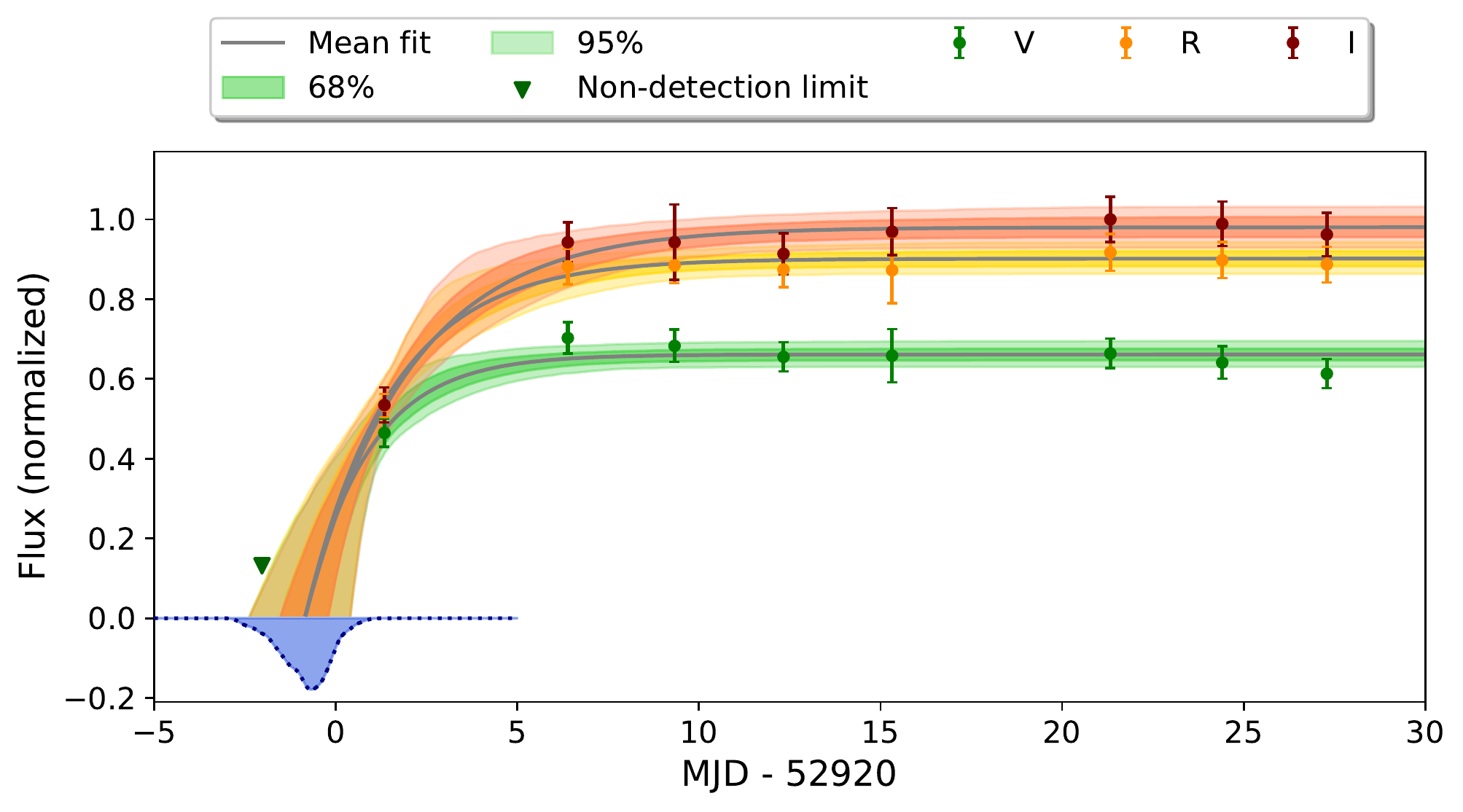}
    \caption{Early light curve fits of SN~2003hl (\textbf{left}) and SN~2003iq (\textbf{right}), and the determined time of explosion posteriors (blue shaded regions).}
    \label{fig:2003hl-2003iq_lc}
\end{figure*}

\begin{figure*}
    \centering
    \includegraphics[width=\linewidth]{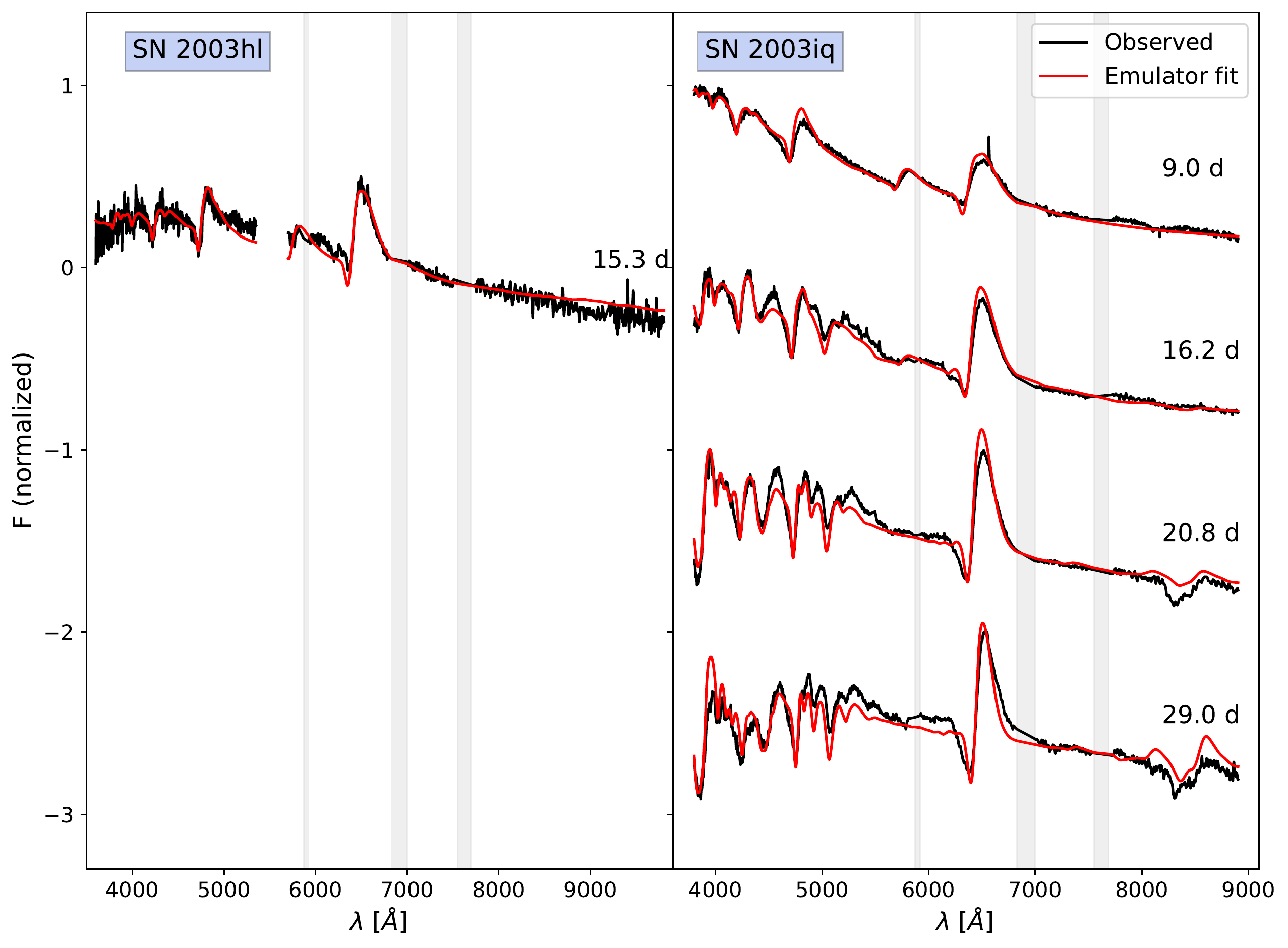}
    \caption{Spectral time series of SNe 2003hl and 2003iq, along with their fits. The grey shaded areas denote telluric regions.}
    \label{fig:NGC772_fits}
\end{figure*}

\begin{figure*}
    \centering
    \includegraphics[width=\linewidth]{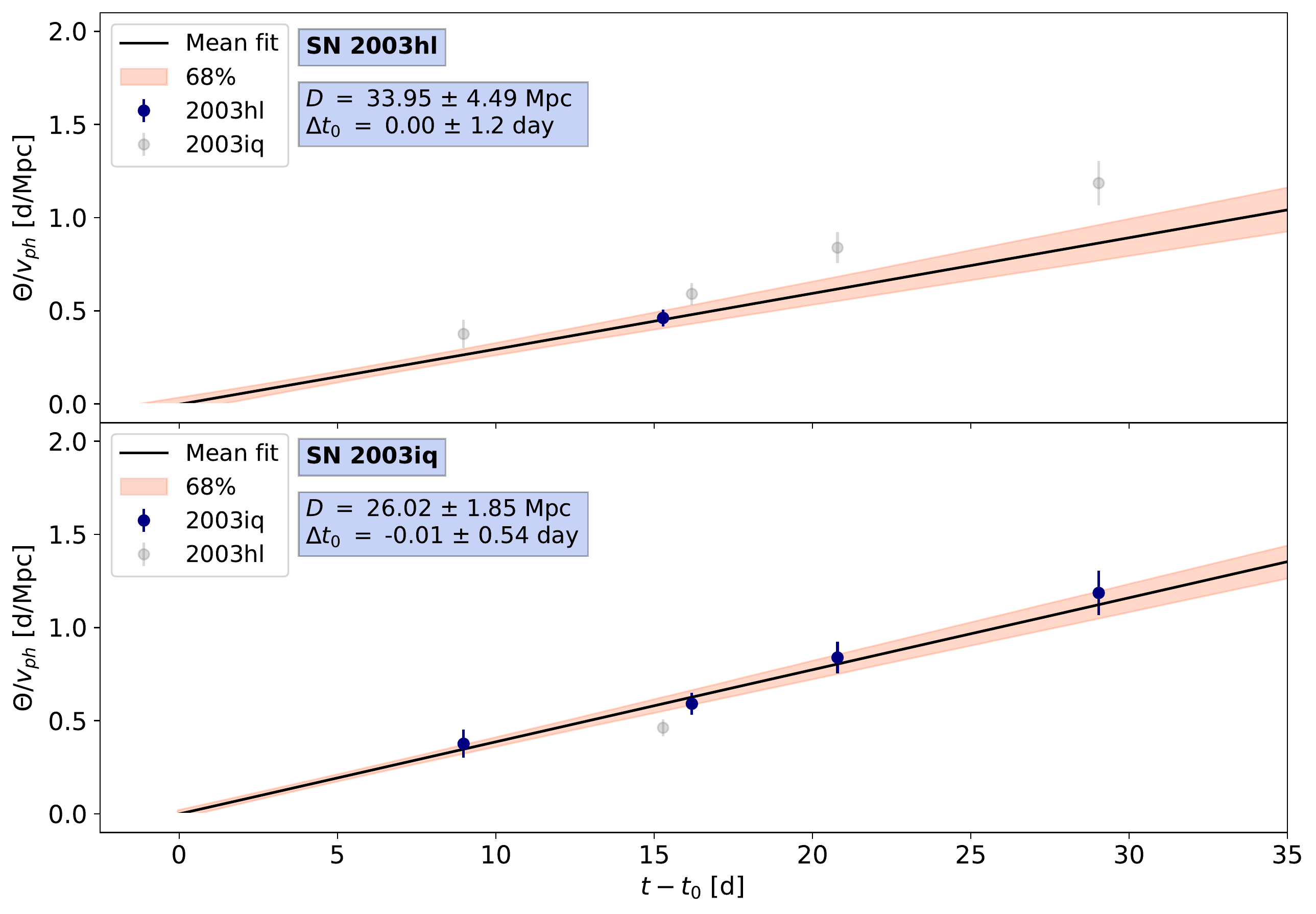}
    \caption{Evolution of $\Theta / v_{\textrm{ph}}$ for SNe 2003hl (\textbf{top} panel) and 2003iq (\textbf{bottom} panel).}
    \label{fig:NGC772_EPM}
\end{figure*}

\subsection{M61}
The spiral galaxy \object{M61} (NGC~4303, Fig.~\ref{fig:hosts}) hosted two Type II supernovae, SN 2008in and 2020jfo. This pair of transients has not been investigated together in the literature yet, since the second supernova occurred only recently.

The first supernova, SN 2008in was discovered on JD 2454827.29 by \cite{2008in}. As M61 was monitored by the Robotic Optical Transient Search Experiment (ROTSE) starting from the week before the discovery \citep{Roy2011}, the fitting of the early light curve allowed for the estimation of the time of explosion independent of the EPM, from which we obtained $t_0 = \textrm{JD } 2454824.51^{+0.19}_{-0.14}$ (Fig.~\ref{fig:2020jfo-2008in_lc}). This supernova was found to belong to the peculiar, subluminous group of SNe IIP \citep{Roy2011}. The spectral sequence for our analysis consisted of spectra obtained by \cite{Roy2011} complemented with more recently published spectra from \cite{Hicken2017}. 

SN 2020jfo was discovered on JD 2458975.70 by the Zwicky Transient Facility (ZTF, \citealt{2020jfo}). Owing to the relatively short cadence of observations of ZTF \citep{ZTF} the time of explosion $t_0$ is not only constrained by a non-detection 4 days pre-discovery, but the supernova was discovered on the rise, hence it was possible to estimate $t_0$ precisely by the fitting of the early light curve (Fig.~\ref{fig:2020jfo-2008in_lc}). We obtained a time of explosion of $t_0 = \textrm{JD } 2458975.37^{+0.10}_{-0.10}$. The spectral time series of this object was presented by \cite{Sollerman2021} and \cite{Teja2022}. For our study, we used seven early time spectra ($t < 20$ days).

The spectral time series and the emulator fits are shown in Fig.~\ref{fig:NGC4303_fits}. For the calculation of the best-fitting models the telluric regions (as marked on the figure) were masked. After fitting, we performed the EPM analysis on both supernovae (Fig.~\ref{fig:NGC4303_EPM}). The light curve fits used for the flux calibration of the spectra are displayed in Appendix ~\ref{Sec:lc_fits}, while the best-fit parameters are listed in Table~\ref{tab:fit_pms}.
The estimated distances to the supernovae are $D = 15.06 \pm 0.71$ Mpc and $D = 14.95 \pm 0.78$ Mpc for SN~2008in and SN~2020jfo respectively. The classical EPM analysis of SN 2008in was conducted previously by \cite{Bose2014}, yielding a distance of $D = 14.51 \pm 1.38$ Mpc, which is consistent with our current estimate. No distance measurements of SN~2020jfo were carried out previously. Although the calibration quality varied significantly from epoch to epoch (since multiple instruments were used for both spectral sequences), the obtained distances are consistent within the uncertainties and agree within 1\% percent.

\begin{figure*}
    \centering
    \includegraphics[width=0.490\linewidth]{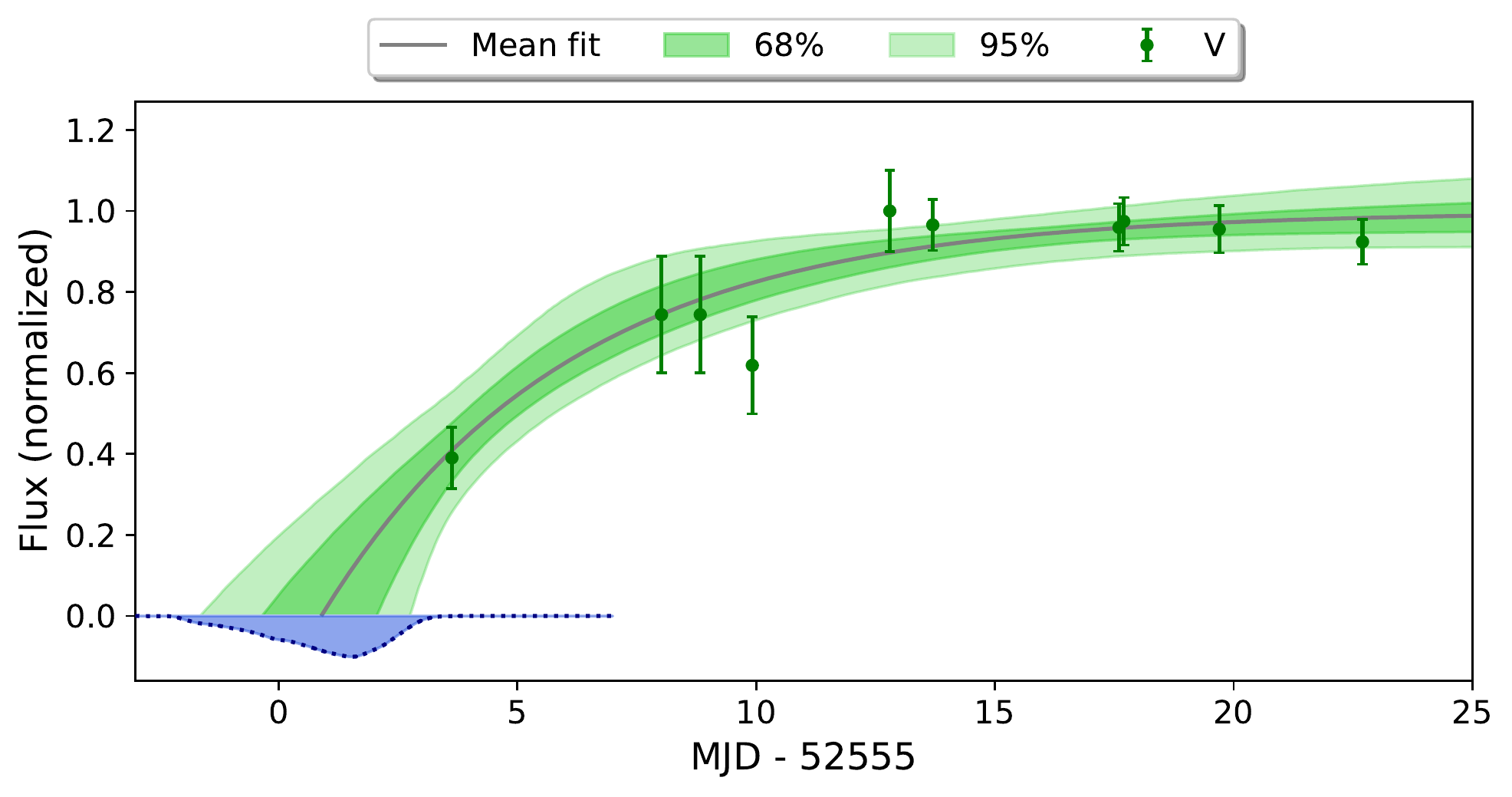}
    \includegraphics[width=0.498\linewidth]{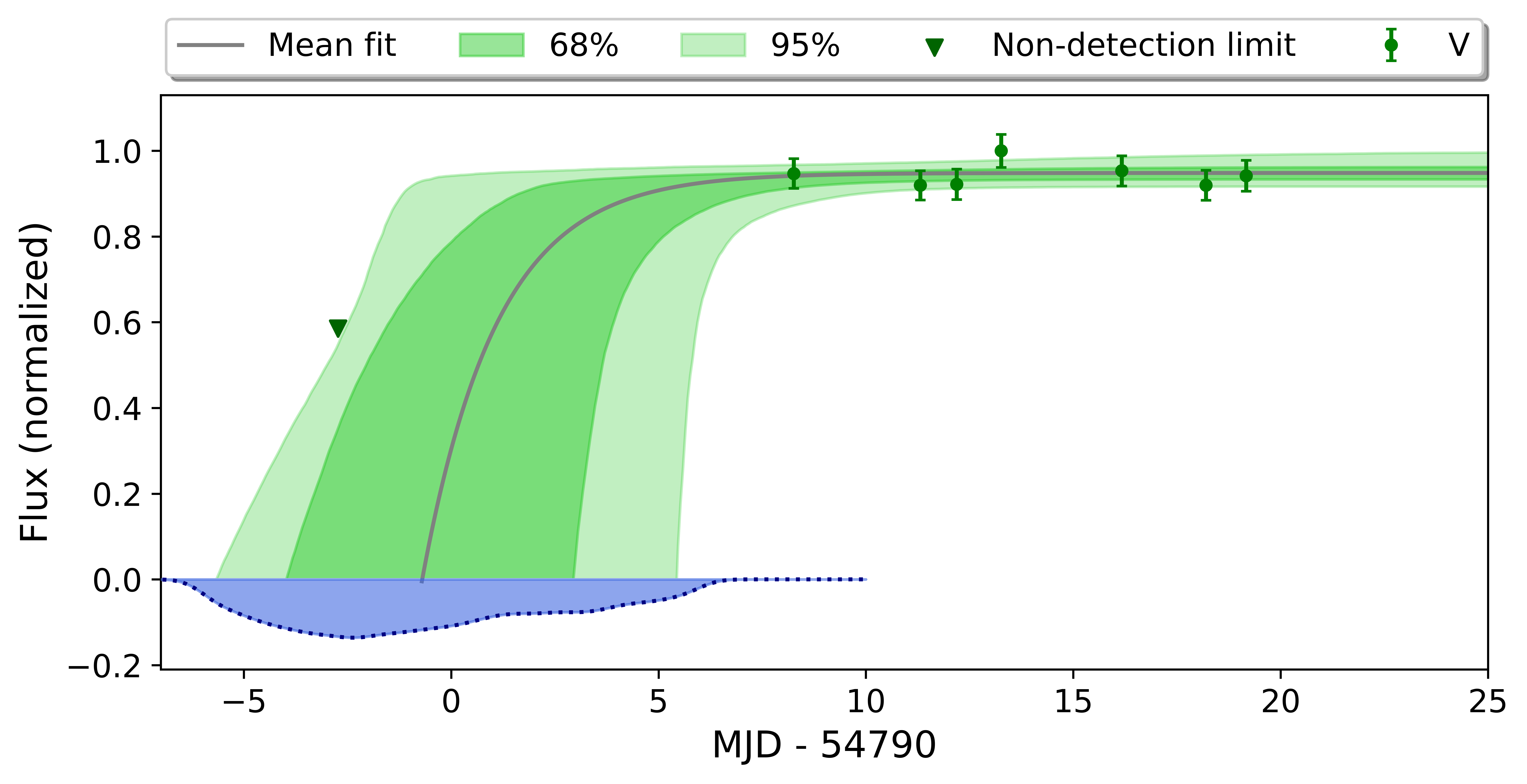}
    \caption{Early light curve fits of SN 2002gw (\textbf{left}) and SN~2008ho (\textbf{right}), and the determined time of explosion posteriors (blue shaded regions).}
    \label{fig:2002gw-2008ho_lc}
\end{figure*}

\begin{figure*}
    \centering
    \includegraphics[width=\linewidth]{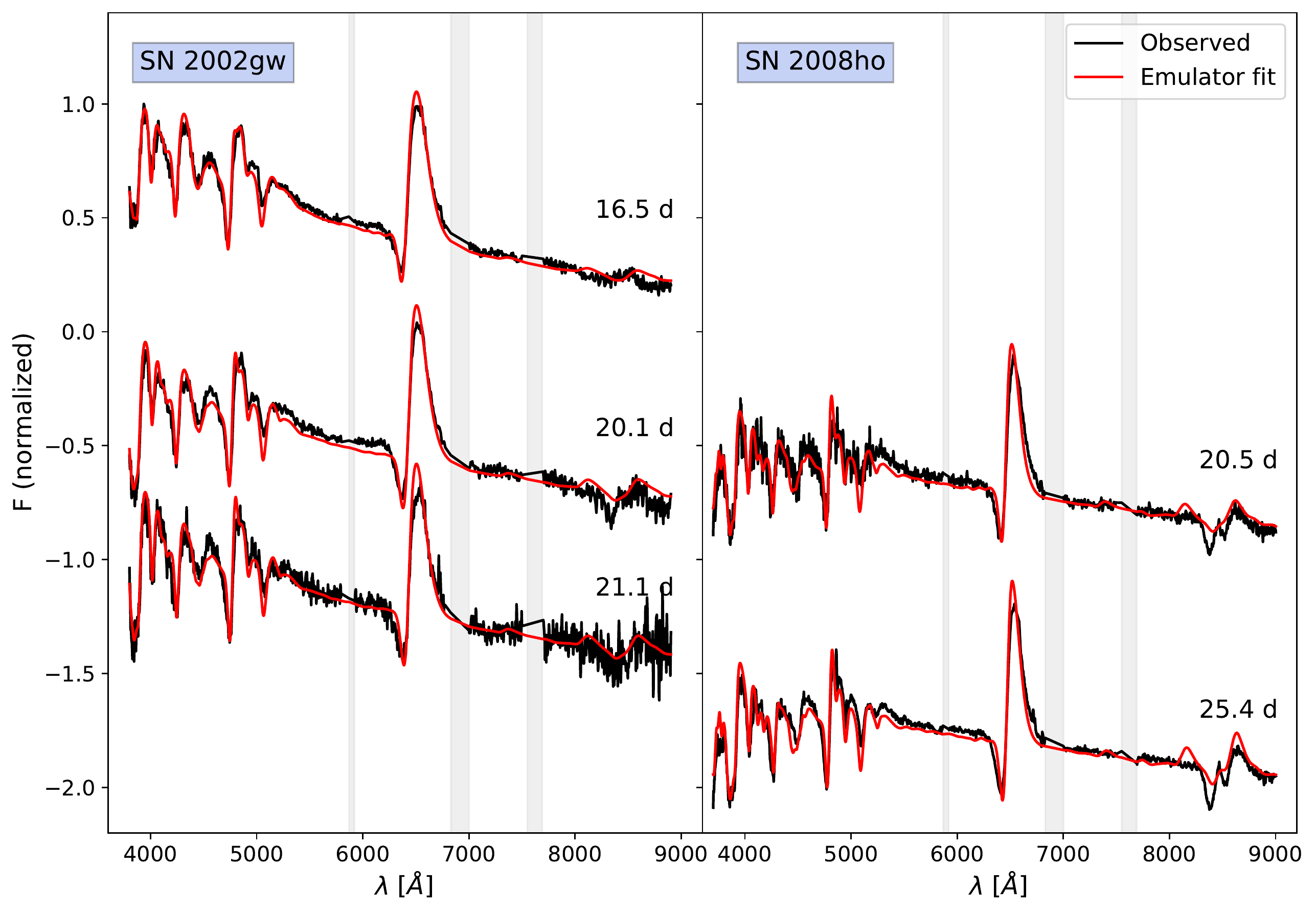}
    \caption{Spectral time series of SNe 2002gw and 2008ho, along with their fits. The grey-shaded areas denote telluric regions.}
    \label{fig:NGC922_fits}
\end{figure*}

\begin{figure*}
    \centering
    \includegraphics[width=\linewidth]{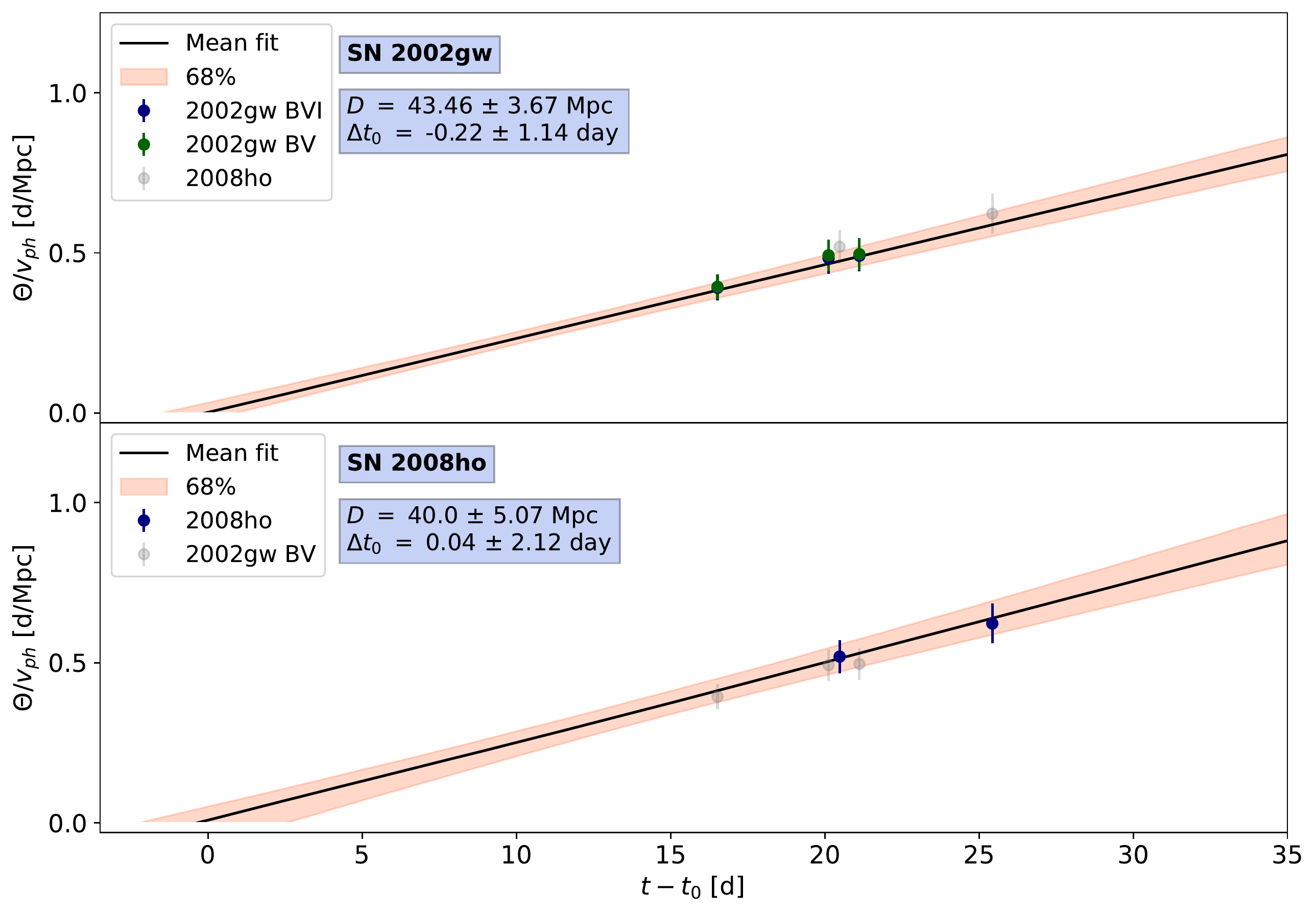}
    \caption{Evolution of $\Theta / v_{\textrm{ph}}$ for SNe 2002gw (\textbf{top} panel) and 2008ho (\textbf{bottom} panel).}
    \label{fig:NGC922_EPM}
\end{figure*}

\begin{figure}
    \centering
    \includegraphics[width=\linewidth]{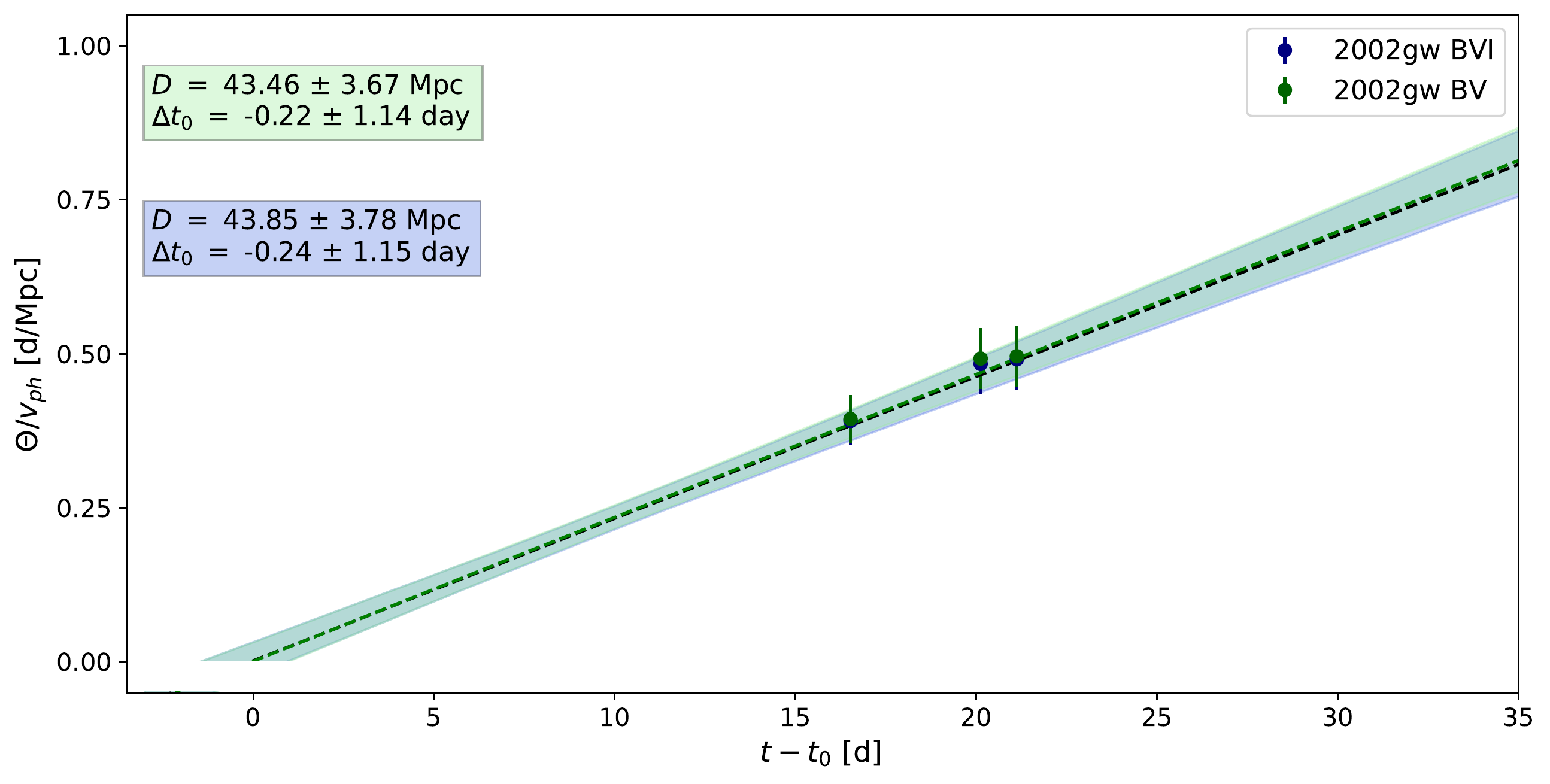}
    \caption{The EPM regression performed for SN 2002gw on the $\Theta / v_{\textrm{ph}}$ values calculated using the $BV$ (green) and $BVI$ (blue) bandpass sets.}
    \label{fig:2002gw_EPM}
\end{figure}

\subsection{NGC~772}
The spiral galaxy \object{NGC~772} (Fig.~\ref{fig:hosts}) is unique, as it not only had been host to two Type II supernovae, but both objects were also observable simultaneously, as they exploded one and a half months after one another. Both supernovae were followed up by the Carnegie Type II Supernova Program (CATS), and their spectral sequences were previously analysed by \cite{Jones2009}.

The first supernova, SN 2003hl, was discovered on JD 2452872.0 by \cite{2003hl}. By applying the method described in Sect.~\ref{sec:Calib} on the early light curve along with the unfiltered pre-discovery KAIT (Katzman Automatic Imaging Telescope, \citealt{Filippenko2001}) non-detection on JD 2452863.0 we estimated the time of explosion to be JD $2452864.62^{+1.18}_{-1.15}$ (Fig.~\ref{fig:2003hl-2003iq_lc}). We assigned the KAIT non-detection to the $V$-band, given that the colour of the supernova is close to zero owing to the combined effect of the very blue spectral energy distribution in such early phases and the reddening, which places the effective wavelength close to the $V$-band even for a red-sensitive CCD. Only one spectrum was obtained for this supernova in the temporal range covered by our emulator. Nevertheless, using the time of explosion, we could still derive an approximate EPM distance.

The second supernova, SN~2003iq was discovered by \cite{2003iq} on JD 2452921.5 during the photometric follow-up observations of SN~2003hl. Owing to the monitoring of the host, a pre-discovery image was taken on JD 2452918.5 by \cite{2003iq} shortly before the first detection, which already constrained the explosion date of SN~2003iq to a range of 3 days. By fitting the early light curve, we estimated its time of explosion to be JD $2452919.71^{+0.59}_{-0.65}$ (Fig.~\ref{fig:2003hl-2003iq_lc}). Apart from the more precisely known $t_0$, this supernova has a more thorough spectral record during the photospheric phase than its sibling, as four spectra were acquired during its early evolution.

The fitted spectral time series of the supernovae are displayed in Fig.~\ref{fig:NGC772_fits}. The EPM regression derived from the obtained physical parameters is shown in Fig.~\ref{fig:NGC772_EPM}. We obtain $D = 33.95 \pm 4.49$ Mpc and $26.02 \pm 1.85$ Mpc for the distance of NGC~772 from SN~2003hl and SN~2003iq respectively. Although we could use only one spectrum for SN~2003hl, and hence the resulting distance estimate is fairly uncertain, both the final result and the $\Theta / v_{\textrm{ph}}$ value around 15\,days show broad consistency with those of SN~2003iq.

By comparing our results with the classical EPM carried out by \cite{Jones2009}, we find significant differences for SN~2003hl and SN~2003iq as well, both for the \cite{E96} and \cite{DH05} correction factors: in the case of 2003hl, our distance is larger than the two \cite{Jones2009} estimates ($D = 17.7 \pm 2.1$ Mpc and $D = 30.3 \pm 6.3$ Mpc for the different correction factors respectively), while for SN~2003iq our distance is shorter than the previous estimates ($D = 36.0 \pm 5.6$ Mpc and $D = 53.3 \pm 17.1$ Mpc, \citealt{Jones2009}). However, the differences between the estimates can be explained by the significantly higher colour excess we obtained for SN~2003iq (with our best estimate being $E(B-V) = 0.14$ mag), and by the fact that we could not make use of more than one spectrum for SN~2003hl, due to the limitations of our modelling approach. On the other hand, by comparing our estimates with previous SCM distances, we find our solution for SN~2003iq is consistent with the previous result of \cite{Poznanski2009} ($D = 26.6 \pm 1.25$ Mpc), while our distance for SN~2003hl is not in tension with the SCM estimate of \cite{Olivares2010} ($D = 25.6 \pm 3.30$ Mpc).

\begin{figure*}
    \centering
    \includegraphics[width=0.495\linewidth]{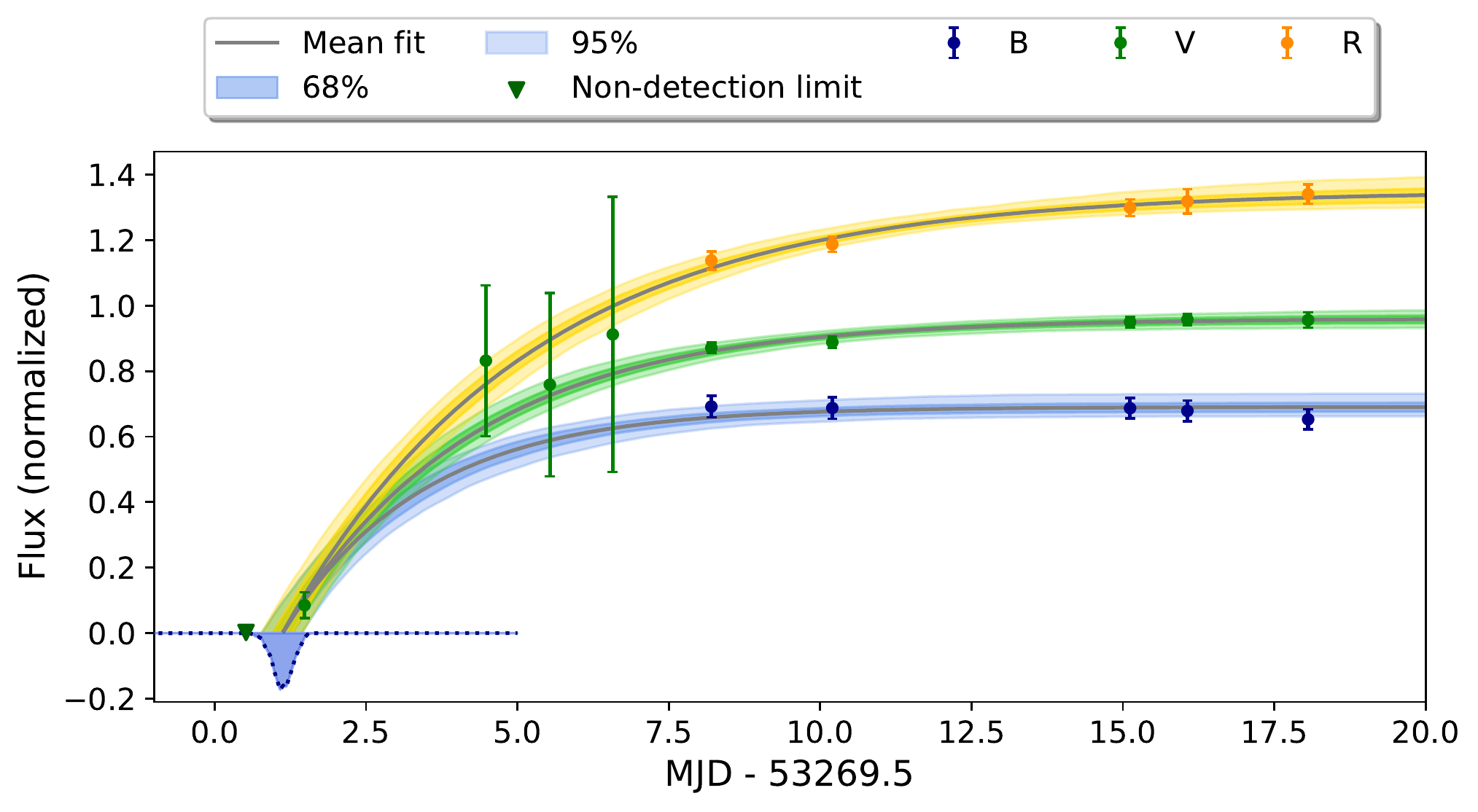}
    \includegraphics[width=0.493\linewidth]{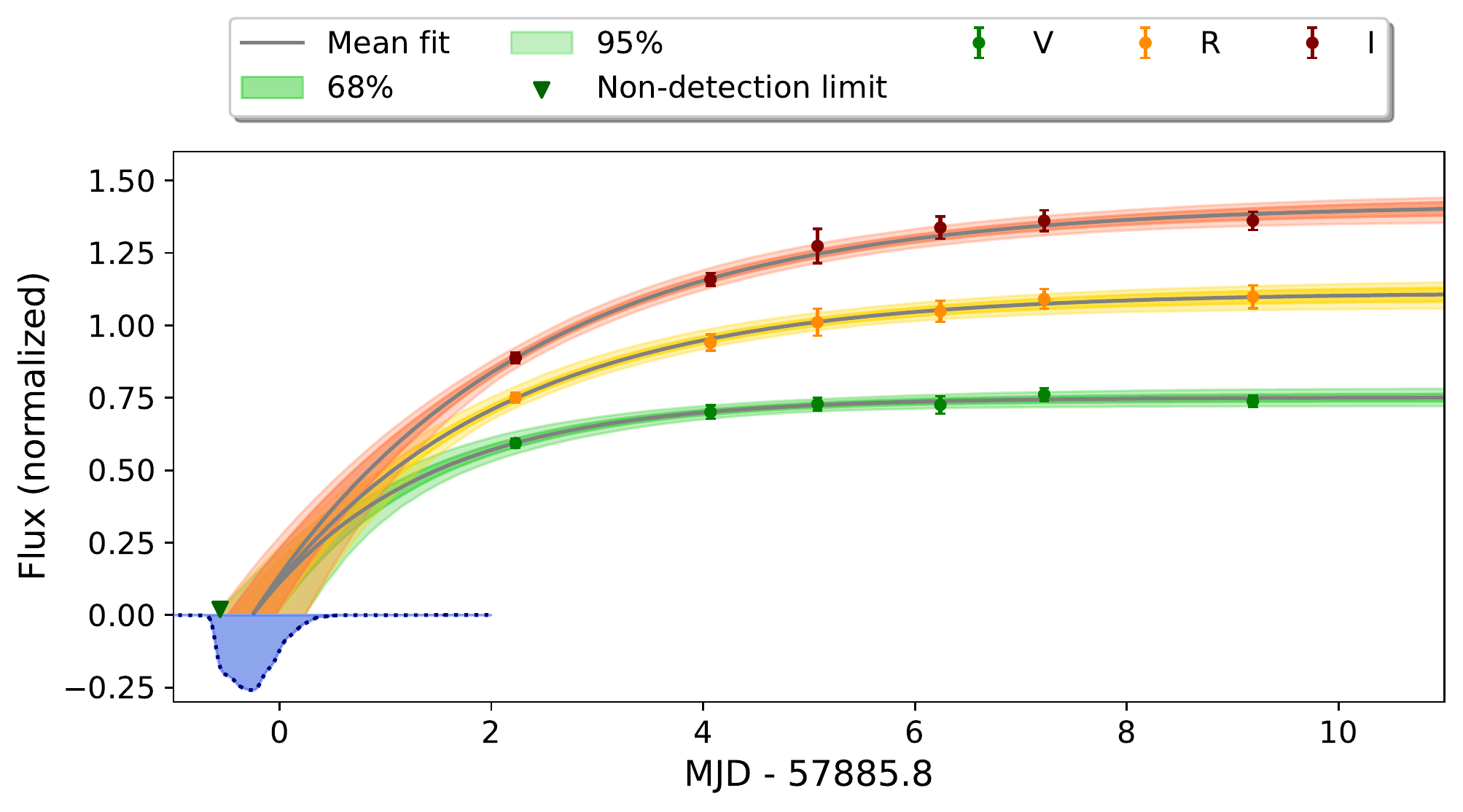}
    \caption{Exponential fit on the early light curves of SN 2004et (\textbf{left}) and SN~2017eaw (\textbf{right}), and the determined time of explosion posteriors (blue shaded regions).}
    \label{fig:2004et-2017eaw_lc}
\end{figure*}

\begin{figure*}
    \centering
    \includegraphics[width=\linewidth]{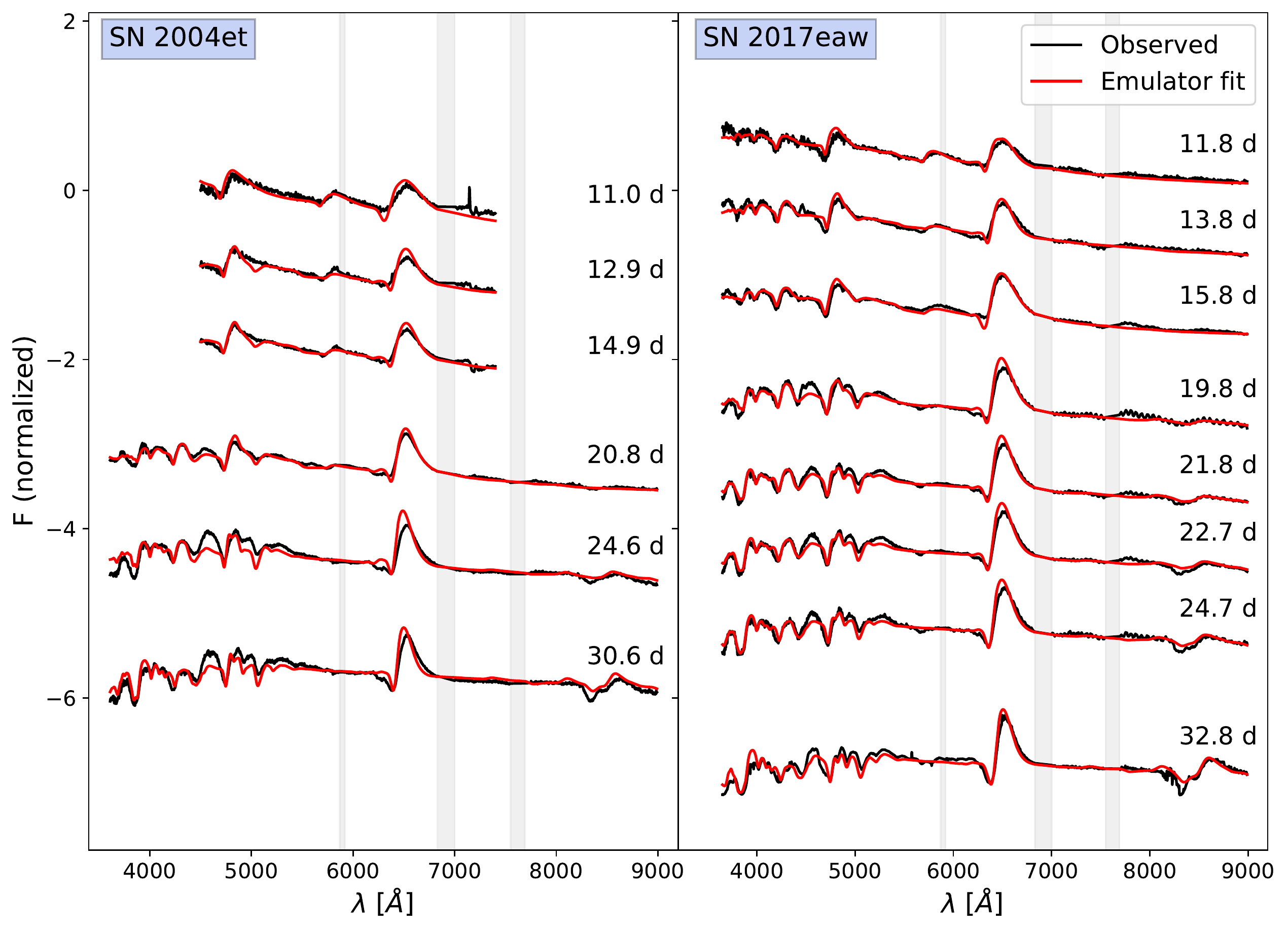}
    \caption{Spectral time series of SNe 2004et and 2017eaw, along with their fits.}
    \label{fig:NGC6946_fits}
\end{figure*}

\begin{figure*}
    \centering
    \includegraphics[width=\linewidth]{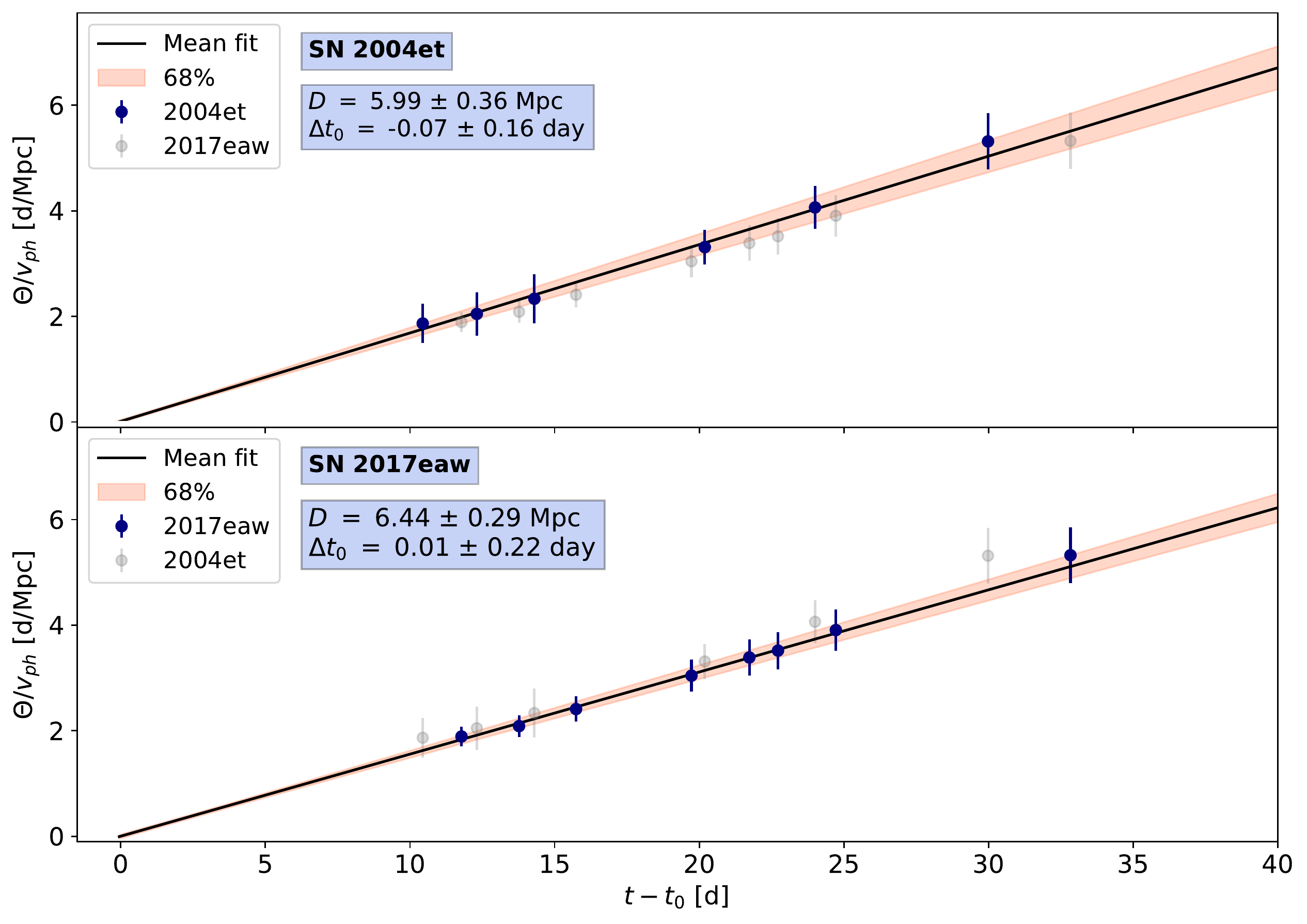}
    \caption{Evolution of $\Theta / v_{\textrm{ph}}$ for SNe 2004et (\textbf{top} panel) and 2017eaw (\textbf{bottom} panel).}
    \label{fig:NGC6946_EPM}
\end{figure*}

\subsection{NGC 922}

The peculiar SBcd type galaxy \object{NGC~922} (Fig.~\ref{fig:hosts}) hosted two Type II supernovae six years apart: SN~2002gw and SN~2008ho. SN~2002gw was discovered on JD 2452560.8 \citep{2002gw}. Although the latest non-detection occurred too far from the discovery to be useful for constraining the explosion date, the fitting of the early light curve (including the unfiltered CCD observations obtained by \citealt{Itagaki2002} and \citealt{2002gw}) gave an estimate of JD $2452556.58^{+0.97}_{-1.40}$ (Fig.~\ref{fig:2002gw-2008ho_lc}), which is consistent with the value obtained through performing the EPM regression in \cite{Jones2009}. In total, three optical spectra were acquired for this supernova by the CATS program \citep{Hamuy2006} in the epoch range that could be used for our EPM analysis.

The second supernova, SN 2008ho, was discovered on JD 2454796.61 by \cite{2008ho}. The last pre-discovery non-detection occurred on JD 2454787.77, which provides a weak constraint on the time of explosion: we found $t_0 =\textrm{JD } 2454789.63^{+3.63}_{-2.54}$ from the light curve fit (Fig.~\ref{fig:2002gw-2008ho_lc}). In total, two spectra were obtained for this supernova through the Carnegie Supernova Project \citep{Hamuy2006}. The spectra were recalibrated using the $BV$ observations obtained through the CSP project (\citetalias{Anderson2022}, private communication). For the EPM, we made use of the $BV$ photometry from the same campaign.


The fitted spectral sequences of SN 2002gw and SN 2008ho are shown in Fig.~\ref{fig:NGC922_fits}. From the EPM regression we obtained a distance for SN~2008ho of $D=$ $40.02 \pm 5.07$ Mpc (Fig.~\ref{fig:NGC922_EPM}). On the other hand, we found $D = 43.46 \pm 3.77$ Mpc for $BV$ bands only and $D = 43.85 \pm 3.78$ Mpc for the full $BVI$ set for SN~2002gw (see Fig.~\ref{fig:2002gw_EPM} for the comparison). Our distance estimates for SN 2002gw fall between the values derived by \cite{Jones2009} for the different dilution factors ($D = 37.4 \pm 4.9$ Mpc and $D = 63.9 \pm 17.0$ Mpc), and agree well with the previous SCM ($D = 48.1 \pm 6.2$ Mpc, \citealt{Olivares2010}) and photospheric magnitude method estimates ($D = 45.10 \pm 3.11$ Mpc, \citealt{Rodriguez2014}). Although the difference is only $\sim 1$\%, we use the $BV$ instead of the $BVI$ distance for SN~2002gw for the plots and the distance consistency check (Sect. \ref{sec:Discuss}) to make the analysis of the siblings as similar as possible. We suspect that the majority of the offset between the distances of the two SNe can be attributed to the relatively large uncertainties on the times of explosion. Nevertheless, the two distances are fully consistent within $1\sigma$.



\subsection{NGC~6946}
\label{sec:6946}
\object{NGC~6946} (Fig.~\ref{fig:hosts}) is a bright face-on SABcd type galaxy, which produced several Type II supernovae: SNe 1948B, 1980K, 2002hh, 2004et and 2017eaw. Due to its proximity, the distance of this galaxy could be estimated through the tip of the red giant branch (TRGB) method \citep{Anand2018} and the planetary nebulae luminosity function (PNLF) relation \citep{Herrmann2008}. Three of the aforementioned supernovae (SNe 2002hh, 2004et and 2017eaw) were observed spectroscopically during the early photospheric phase, which makes it possible to measure their distances through the tailored-EPM. However, only SNe 2004et and 2017eaw were optimal for our purposes: even though the spectral time series and the photometric coverage would have allowed the analysis of SN 2002hh, it was found that this supernova exhibited very strong, two-component reddening (one component arising from the joint effect of the interstellar dust in the host and the Milky Way, the other from large quantities of local dust, see \citealt{2005ApJ...627L.113B}, \citealt{Pozzo2006} and \citealt{2007ApJ...669..525W}). With the currently available data and methodology it is not possible to obtain an accurate extinction correction. Considering the very high reddening, any error in our extinction correction would significantly impact the inferred distance. Hence we chose to exclude SN~2002hh from the analysis.

SN~2004et was discovered on JD 2453273 by \cite{2004et}, and extensively followed up by the 2-m Himalayan Chandra Telescope (HCT) of the Indian Astronomical Observatory (IAO) and the 3-m Shane telescope at the Lick Observatory. The resulting spectral times series has been previously studied by \cite{Sahu2006}. However, the spectra taken with the HCT were subject to calibration issues, which could not be corrected with our standard linear flux re-calibration. Since these issues can influence the fitting significantly, we attempted to empirically correct them, using spectra taken with the Keck and Lick telescopes. This procedure is detailed in Appendix ~\ref{app:2004et_corr}. We completed this spectral time series with three additional early-time spectra obtained at the David Dunlap Observatory \citep[ courtesy of J\'{o}zsef Vink\'{o}]{TakatsVinko2012}. However, since these spectra covered only a very narrow wavelength range, we could not apply the re-calibration procedure from Sect.~\ref{sec:flux_calibration}. The pre-discovery non-detection of the supernova and the subsequent early photometry from \cite{Li2004} allowed for an accurate determination of the time of explosion through the fitting of the early light curve (Fig.~\ref{fig:2004et-2017eaw_lc}), which yielded $t_0 = \textrm{JD } 2453271.19^{+0.16}_{-0.17}$. Along with the tight constraints on the time of explosion, the six spectra obtained in the first month after explosion also allowed for a precise EPM analysis. 

SN~2017eaw, was discovered by \cite{2017eaw} on JD 2457885.78, and was extensively followed up with spectroscopic observations at the Las Cumbres Observatory 1~m telescope and at the McDonald Observatory with the Low Resolution Spectrograph 2 mounted on the 10 m Hobby–Eberly Telescope. The observations are described in \cite{Szalai2019}. The supernova was discovered early, on the rise, and a non-detection is also available from a pre-discovery observation of the host, which yielded a time of explosion of $t_0 = \textrm{JD } 2457886.01^{+0.25}_{-0.20}$. In terms of spectroscopy, this supernova had the best-sampled and most homogenous spectral sequence in our sample, resulting in eight well-calibrated spectra.

The calibrated and fitted spectral sequences are shown in Fig.~\ref{fig:NGC6946_fits}. For SN~2004et, the EPM regression yielded a distance of $D = 5.99 \pm 0.36$ Mpc, while for SN~2017eaw the resulting value was $D = 6.44 \pm 0.29$ Mpc (Fig.~\ref{fig:NGC6946_EPM}). The distance for SN~2004et is 20\% higher than the classical EPM value inferred by \cite{TakatsVinko2012} and \cite{Szalai2019}, while for SN~2017eaw the value we obtained is about 5-20\% lower than the value from \cite{Szalai2019} (depending on the assumed reddening and the chosen epoch range of spectra of the reference work, see \citealt{Szalai2019} for details). We note that our distance estimates are in much better agreement with one another, which highlights the advantage of the tailored-EPM over the classical EPM approach. The distances are also consistent with the various independent distances obtained for NGC~6946: apart from the previous EPM-based values, our estimates are in agreement with SCM distances of \cite{Poznanski2009} and \cite{deJaeger2017} (both of which yielded $D = 6.69 \pm 0.30$ Mpc for SN~2004et) and the PNLF distance (\citealt{Herrmann2008}, $D = 6.1 \pm 0.6$ Mpc), but is slightly lower than the latest TRGB estimate ($D = 6.95 \pm 0.20$ Mpc, \citealt{Anand2021}, from the Extragalactic Distance Database\footnote{\url{https://edd.ifa.hawaii.edu/}}).

\begin{table}[]
    \centering
    \resizebox{\columnwidth}{!}{
    \begin{tabular}{l l c c c}
    Host & SN & $D$ [Mpc] & $t_0$ (MJD) & $E(B-V)$ [mag] \\
    \hline
    M61 & 2008in & $15.06(0.71)$ & $54824.15(0.15)$ & 0.06(0.03)\\
        & 2020jfo & $14.95(0.78)$ & $58975.08(0.34)$ & 0.07(0.04)\\
    \hline
    NGC~722 & 2003hl & $33.95(4.49)$ & $52864.26(1.20)$ & 0.31*\\
            & 2003iq & $26.02(1.85)$ & $52919.20(0.54)$ & 0.15(0.05)\\
    \hline
    NGC~922 & 2002gw & $43.46(3.67)$ & $52555.84(1.15)$ & 0.12(0.03)\\
            & 2008ho & $40.02(5.07)$ & $54789.17(2.12)$ & 0.21*\\
    \hline
    NGC~6946 & 2004et & $5.99(0.36)$ & $53270.62(0.16)$ & 0.38(0.10)\\
             & 2017eaw & $6.44(0.29)$ & $57885.52(0.22)$ & 0.35(0.04)\\
    \hline
    \end{tabular}
    }
    \caption{Summary table of the EPM results. $t_0$ denotes the time of explosion as obtained by the EPM regression. The $E(B-V)$ values listed here are those determined by the spectral fitting. The values in brackets denote the uncertainties of the given quantities. In the starred (*) cases we either had only one spectrum to fit, or both spectra yielded the same reddening value, which did not allow us to give a reasonable empirical uncertainty.}
    \label{tab:EPM_results}
\end{table}

\begin{table*}
    \centering
    \begin{tabular}{ l  c c c c c c}
    Galaxy & SN 1 & D$_{\textrm{SN\,1}}$ & SN 2 & D$_{\textrm{SN\,2}}$  & Bayes factor & Prior\\
    \hline
    NGC~772 & 2003hl & $33.95 \pm 4.49$ Mpc & 2003iq & $26.02 \pm 1.85$ Mpc & 0.890 & 13.44 -- 53.77 Mpc\\
    NGC~922 & 2002gw & $43.46 \pm 3.67$ Mpc & 2008ho & $40.02 \pm 5.07$ Mpc & 3.448 & 21.29 -- 85.16 Mpc\\
    NGC~4303 & 2008in & $15.06 \pm 0.71$ Mpc & 2020jfo & $14.95 \pm 0.78$ Mpc & 7.944 & 7.31 -- 29.25 Mpc\\
    NGC~6946 & 2004et & $5.99 \pm 0.36$ Mpc & 2017eaw & $6.44 \pm 0.29$ Mpc & 4.272 & 2.77 -- 11.09 Mpc\\
    \hline
    \end{tabular}
    \caption{Bayes factors for the comparisons of the distance posteriors of the various SN pairs.}
    \label{tab:Bayes}
\end{table*}

\section{Discussion}
\label{sec:Discuss}
As shown in the previous sections, the fitting procedure not only yields distances with a claimed uncertainty of $\sim$10\% or better (Table~\ref{tab:EPM_results}), it does so by requiring only a limited amount of modelling choices. Due to various uncertainties in the observations and the modelling, these distances can be slightly different for the supernova siblings. The main question, in this case, is whether the estimated distances of the siblings are consistent with one another. Assuming an average galaxy, a reliable upper limit for the thickness of the disc is 10~kpc, while the width of the disc is on the order of ~100~kpc \citep{Gilmore1983,Zanisi2020}; consequently, in an ideal case, assuming face-on galaxies, the distances inferred for the supernova siblings should match to an uncertainty of $\pm0.01$~Mpc, which corresponds to only $\sim 0.2\%$ even for the most nearby pair. For more inclined galaxies the uncertainties can be one order of magnitude larger; however, since the closest hosts we investigated are very likely low inclination galaxies (M~61 and NGC~6946), and the farther away ones are not viewed edge-on either (NGC~722 and NGC~922), the offset between the siblings should remain sub-1\% relative to the distance of the galaxy. Hence, the distance estimation to siblings provides an empirical test to assess the effect of these uncertainties and systematics (e.g. the CSM interaction, or data calibration issues, among others). It allows us to test, whether our results are not only \emph{precise}, but \emph{accurate}.

To test the consistency of the obtained distances, we performed Bayesian model comparison. We adopted the methodology used by \cite{Wong2020} for assessing whether pairs of strongly lensed quasars favour a single global set of cosmological parameters or individual cosmologies for each lens that are inconsistent with one another. We applied this method to the distance estimates of the siblings: we thus asked the question whether the measured distance posteriors of a pair have more likely been generated from a single underlying distance $D^{\textrm{gal}}$ or from two distinct distances $D^{\textrm{ind,1}}$ and $D^{\textrm{ind,2}}$ (and are thus inconsistent). The probability ratio of the two scenarios is called the Bayes factor and can be calculated as

\begin{equation}
    F_{ij} = \frac{P(\mathbf{d_{i}},\mathbf{d_{j}}|D^{\textrm{gal}})}{P(\mathbf{d_{i}}|D^{\textrm{ind,i}})P(\mathbf{d_{j}}|D^{\textrm{ind,j}})} = \frac{\int^{D_2}_{D_1} \mathbf{d_{i}}\mathbf{d_{j}} P(D) dD}{\int^{D_2}_{D_1} \mathbf{d_{i}} P(D) dD \int^{D_2}_{D_1} \mathbf{d_{j}} P(D) dD },
\end{equation}
where $\mathbf{d_{i}}$ and $\mathbf{d_{j}}$ denote the distance posteriors of the individual supernovae in each pair. We chose a uniform prior $P(D)$ around the average distance of the host (based on previous measurements, as quoted on NASA/IPAC Extragalactic Database, NED\footnote{\url{https://ned.ipac.caltech.edu/}}), ranging from half to twice that distance. The quoted Bayes factors can thus be understood as lower limits since they scale linearly with the width of the distance prior. If the obtained Bayes factor exceeds unity, this can be interpreted as a sign of consistency between the two posterior distributions in accordance with the table of \cite{Kass1995}. The higher the $F$ value, the stronger the consistency between the siblings.

\begin{figure}
    \centering
    \includegraphics[width=0.98\linewidth]{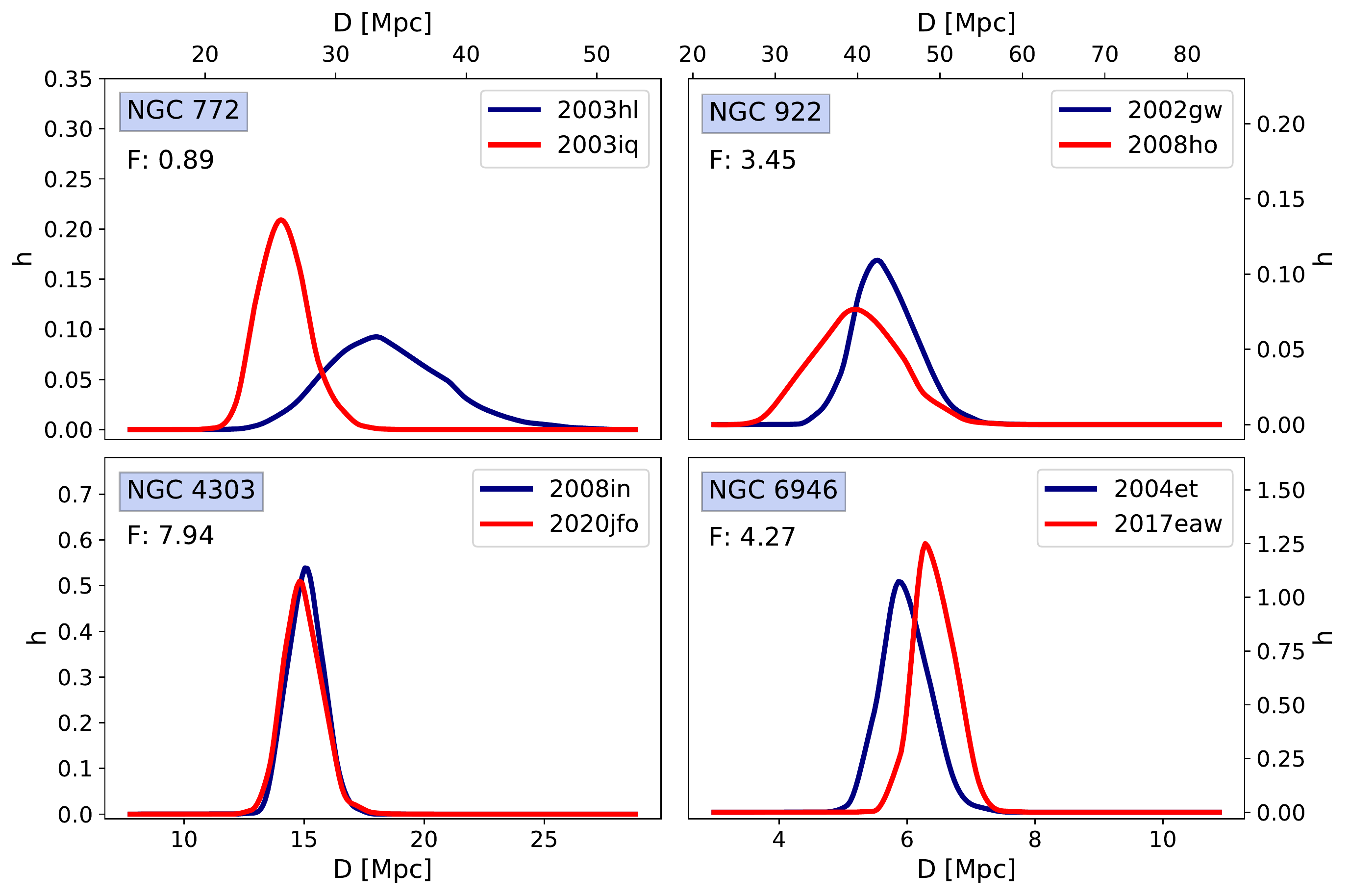}
    \caption{Distance posteriors obtained from the EPM analysis of the supernovae for the various host galaxies. The calculated Bayes factors are displayed in the top-left corners of the panels.}
    \label{fig:posteriors}
\end{figure}

The distance posteriors from the EPM analysis are shown in Fig.~\ref{fig:posteriors}, while the obtained Bayes factors are listed in Table~\ref{tab:Bayes}. The consistency of the distance posteriors for the various supernovae is good, with Bayes factors being close to or over 5 for two cases, showing remarkable agreement. In the case of NGC~772 and NGC~922 the values of the Bayes factors are lower, which might be attributed to the lack of a well-sampled spectral sequence for SN~2003hl, and the lack of well-constrained explosion date for SNe~2002gw and 2008ho. However, the remaining cases show that the fitting method and the analysis result in distances that are not only consistent for supernovae in the same host but are also precise to a degree of a few per cent. The analysis showed as well that some of the greatest limiting factors of the method are an insufficient amount of spectra, poor quality and calibration-level of the obtained spectral times series and weak constraints on the time of explosion. If instead all the conditions for the above are met, the EPM analysis can yield highly precise and consistent values for the supernova pairs, as in the case of SNe~2004et and 2017eaw or in the case of SNe~2008in and 2020jfo. 

We offer the caveat that some of our results may be influenced by the choice of $R_V = 3.1$. If the true $R_V$ is different, this could in principle cause notable offsets between the sibling distances, provided the differential reddening between the objects is non-negligible and the measurement uncertainties are low enough. In our sample, however, the pairs with the most significant differential reddening had also the largest uncertainty on their distance and vice versa. Hence it is currently not possible to estimate the scale of this effect on the basis of our four sibling pairs.

\section{Summary}
\label{sec:Summary}
In this work, we investigated the consistency of the EPM distances of sibling Type II SNe, calculated based on the spectral fitting method introduced by \cite{Vogl2020}. According to our analysis, the method not only yields precise results with an estimated individual distance uncertainty that is better than $\sim$10\%, despite using literature data from a wide range of sources, but the resulting distances are consistent for SNe that are within the same host galaxy. The degree of consistency depends on how well the supernovae were observed, with the worst results occurring when the quality of the data barely allows for the EPM analysis, while the best consistency is achieved when both supernovae are similarly well observed and have a good constraint on the time of explosion. In cases of M61 and NGC~6946 highly consistent distances are derived with mismatches of less than 5\%.

The high level of consistency between siblings also shows that if there are any other systematics present between the various SNe, they are likely subdominant compared to the effect of observation quality. Such systematics may include different reddening, CSM interaction, explosion energetics and metallicities. Checking for the effect and scale of such systematics will require a larger set of siblings with data quality matching that of the siblings of M~61 or NGC~6946.

Apart from checking for the internal consistency of the method, we obtained precise distances for the four investigated host galaxies, even though we used literature data, with a highly varying level of calibration. Furthermore, the tailored-EPM provides absolute distances, is physics-based, is affected by different systematics than the other existing distance estimation techniques and can be utilized completely independently of them. These properties and the presented analysis in this paper demonstrate the potential of the tailored-EPM to provide not only precise distances but a precise Hubble parameter as well. Such an analysis is currently being conducted on high-quality spectral time series obtained for SNe II in the Hubble Flow by the \texttt{adH0cc} collaboration (accurate determination of H0 with core-collapse supernovae\footnote{\url{https://adh0cc.github.io/}}), which will provide important clues to the Hubble tension.

\section{Contributor roles\protect\footnotemark}
\footnotetext{Using the CRT standard (see \url{https://credit.niso.org/})}
\stepcounter{footnote}
\begin{enumerate}
    \item Conceptualization: B.~Leibundgut, W.~Hillebrandt, C.~Vogl
    \item Data curation: G.~Cs\"ornyei
    \item Formal analysis: G.~Cs\"ornyei
    \item Funding acquisition: W.~Hillebrandt
    \item Investigation: G.~Cs\"ornyei, C.~Vogl, S.~Taubenberger, A.~Fl\"ors
    \item Methodology: B.~Leibundgut, W.~Hillebrandt, C.~Vogl
    \item Project administration: B.~Leibundgut, W.~Hillebrandt, C.~Vogl
    \item Resources: W.~Hillebrandt
    \item Software: G.~Cs\"ornyei, C.~Vogl
    \item Supervision: B.~Leibundgut, W.~Hillebrandt, C.~Vogl
    \item Validation: G.~Cs\"ornyei
    \item Visualization: G.~Cs\"ornyei
    \item Writing -- original draft: G.~Cs\"ornyei, C.~Vogl
    \item Writing -- review and editing: B.~Leibundgut, W.~Hillebrandt, S.~Taubenberger, A.~Fl\"ors, S.~Blondin, M.~G.~Cudmani, A.~Holas, S.~Kressierer
\end{enumerate}

\section*{Acknowledgements}
We specially thank J. Anderson, M. Stritzinger and the CSP collaboration for providing us pre-published data, which helped us to improve our analysis. The research was completed with the extensive use of Python, along with the \texttt{numpy} \citep{numpy}, \texttt{scipy} \citep{scipy} and \texttt{astropy} \citep{astropy} modules. This research made use of \textsc{Tardis}, a community-developed software package for spectral synthesis in supernovae \citep{TARDIS, TARDIS2}. The development of \textsc{Tardis} received support from the Google Summer of Code initiative and from ESA's Summer of Code in Space program. \textsc{Tardis} makes extensive use of Astropy and PyNE. Based on observations obtained with the Samuel Oschin 48-inch Telescope at the Palomar Observatory as part of the Zwicky Transient Facility project. ZTF is supported by the National Science Foundation under Grant No. AST-1440341 and a collaboration including Caltech, IPAC, the Weizmann Institute for Science, the Oskar Klein Center at Stockholm University, the University of Maryland, the University of Washington, Deutsches Elektronen-Synchrotron and Humboldt University, Los Alamos National Laboratories, the TANGO Consortium of Taiwan, the University of Wisconsin at Milwaukee, and Lawrence Berkeley National Laboratories. Operations are conducted by COO, IPAC, and UW. This project utilizes data obtained by the Robotic Optical Transient Search Experiment.  ROTSE is a collaboration of Lawrence Livermore National Lab, Los Alamos National Lab, and the University of Michigan (\url{www.umich.edu/~rotse}). CV and WH were supported for part of this work by the Excellence Cluster ORIGINS, which is funded by the Deutsche Forschungsgemeinschaft (DFG, German Research Foundation) under Germany's Excellence Strategy-EXC-2094-390783311. The work of AH was supported by the Deutsche Forschungsgemeinschaft (DFG, German Research Foundation) – Project-ID 138713538 – SFB 881 ("The Milky Way System", subproject A10). AH acknowledges support by the Klaus Tschira Foundation. AH is a Fellow of the International Max Planck Research School for Astronomy and Cosmic Physics at the University of Heidelberg (IMPRS-HD). This work was supported by the 'Programme National de Physique Stellaire' (PNPS) of CNRS/INSU co-funded by CEA and CNES. SB acknowledges support from the ESO Scientific Visitor Programme in Garching. AF acknowledges support by the European Research Council (ERC) under the European Union’s Horizon 2020 research and innovation program (ERC Advanced Grant KILONOVA No. 885281). ST has received funding from the European Research Council (ERC) under the European Union’s Horizon 2020 research and innovation program (LENSNOVA: grant agreement No 771776).

\bibliographystyle{aa}
\bibliography{main_sibling_arxiv}

\begin{appendix}


\section{Correction of the SN~2004et HCT spectra}
\label{app:2004et_corr}
\begin{figure}
    \centering
    \includegraphics[width=0.98\linewidth]{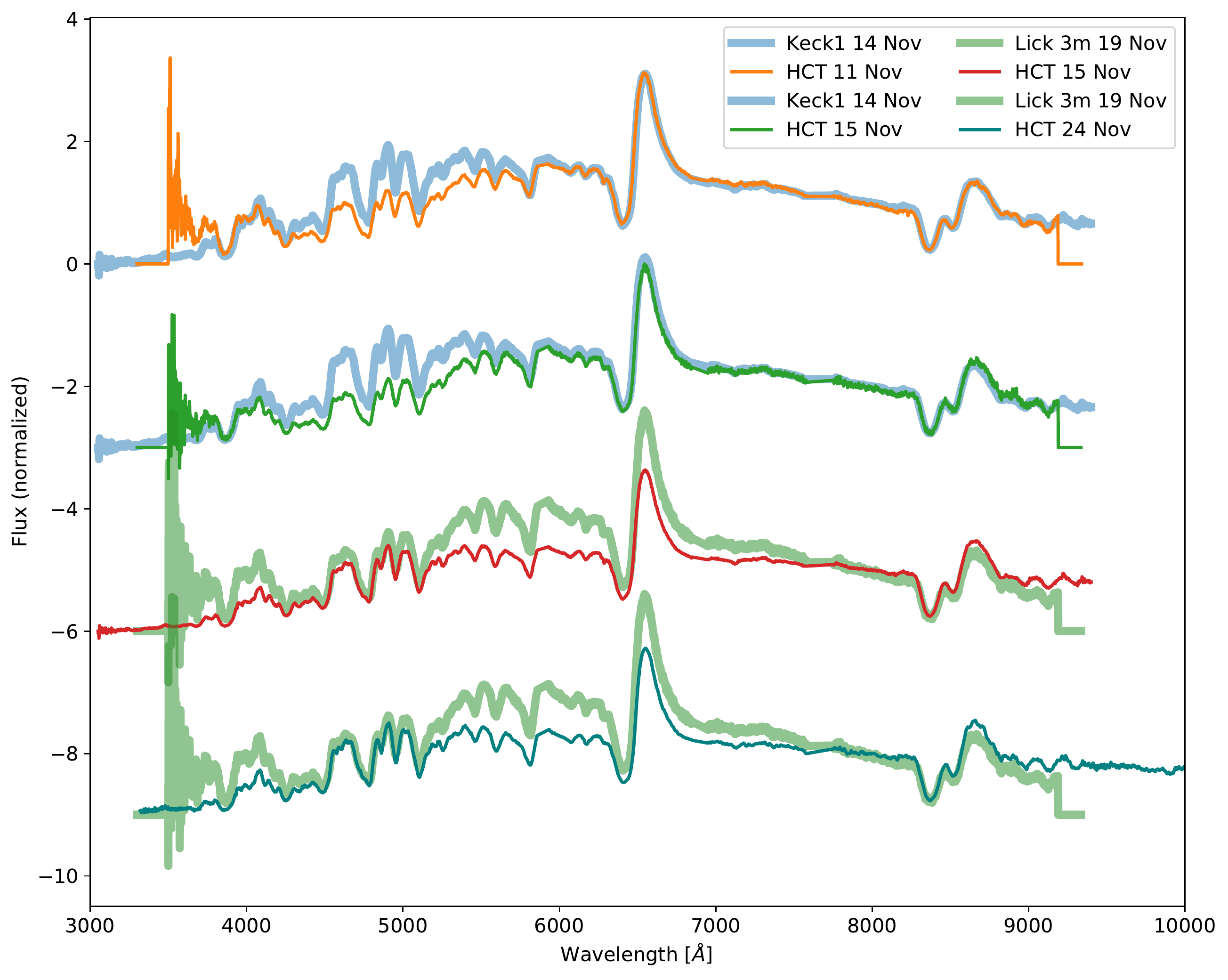}
    \caption{Comparison of the spectra of SN~2004et taken at the HCT and the Lick observatory. The discrepancy between the two spectral sets below the 7000 $\AA$ is clear.}
    \label{fig:2004et_issue}
\end{figure}

By comparing the spectra of SN 2004et taken at the HCT and Lick Observatory at similar epochs, we found that the former show a significant flux deficit in the wavelength range covered mainly by the Bessell $V$ and $R$ bands as shown in the Fig.~\ref{fig:2004et_issue}. This trend could not be corrected by a linear flux correction based on the photometry.

\begin{figure}
    \centering
    \includegraphics[width=0.98\linewidth]{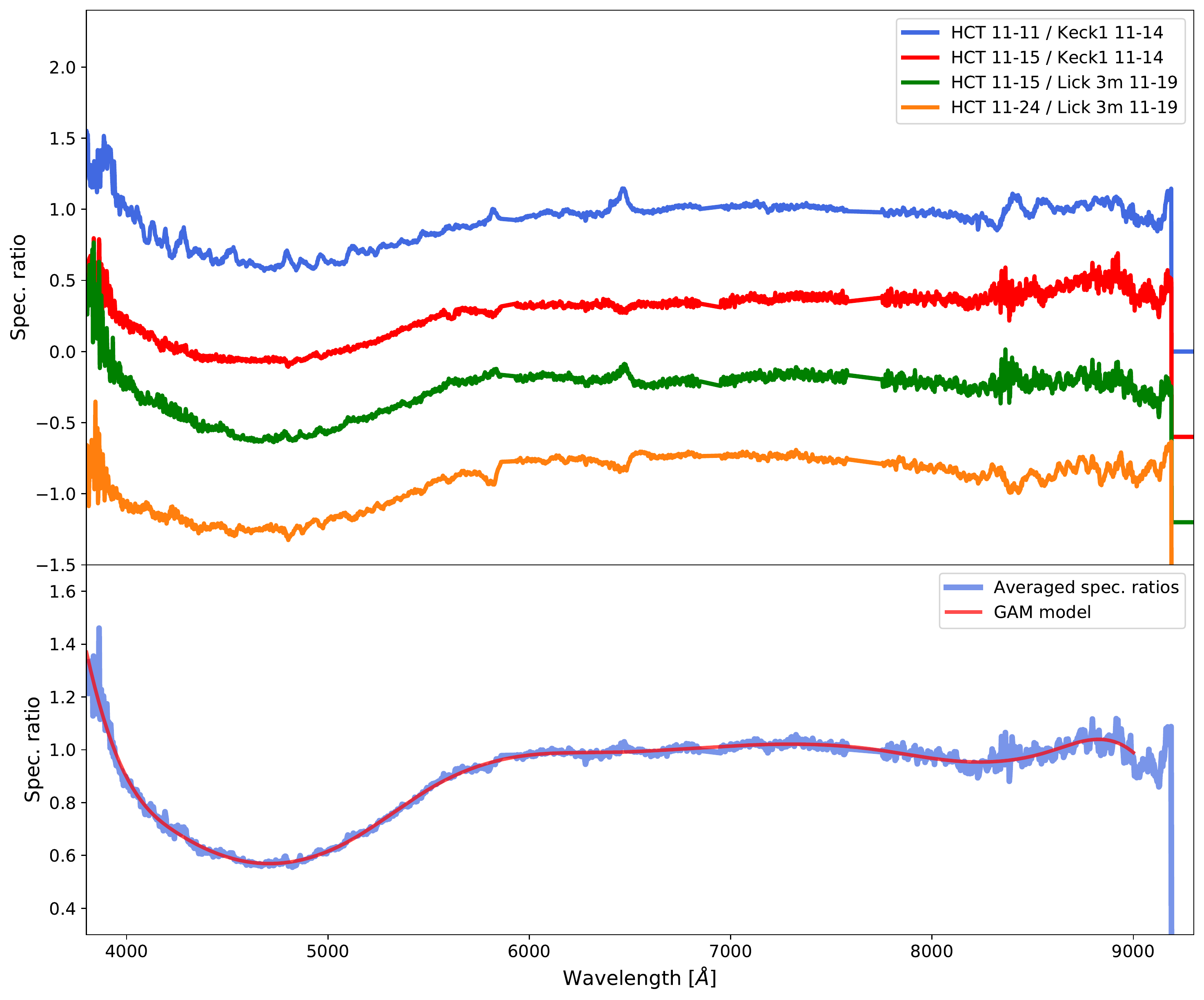}
    \caption{\textbf{Top panel}: the ratios of selected HCT and Keck/Lick spectra that are close in time. \textbf{Bottom panel}: the average curve calculated from the four individual ones on the top panel, along  with its GAM fit.}
    \label{fig:2004et_fix}
\end{figure}

\begin{figure}
    \centering
    \includegraphics[width=0.98\linewidth]{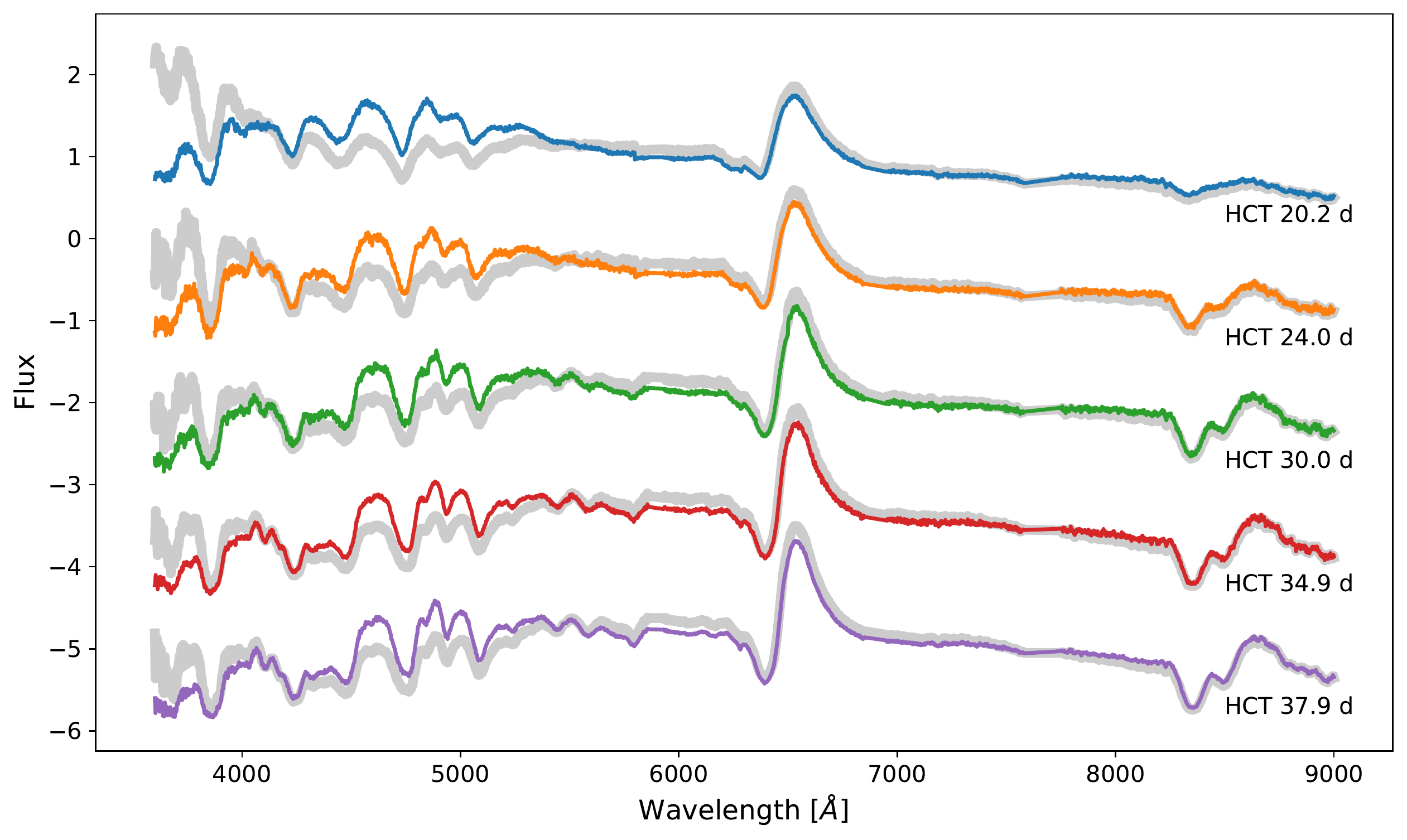}
    \caption{The spectral sequence of SN~2004et obtained by HCT. The grey curves show the uncorrected spectra, while the coloured ones show those for which we applied the non-linear correction derived from the Keck/Lick spectra.}
    \label{fig:2004et_clear}
\end{figure}

To fix this issue, we took the ratios of the uncorrected HCT and the re-calibrated Keck and Lick spectra taken at close epochs (which otherwise are too old for the emulator-based fitting, see top panel of Fig. ~\ref{fig:2004et_fix}), then averaged the obtained trend, assuming that it is the same for every HCT spectrum, and fitted it using Generalized Additive Models (GAMs, \citealt{GAM}, by applying the \texttt{pyGAM} package from Python, \citealt{pygam}), as shown in the bottom panel of Fig.~\ref{fig:2004et_fix}. We then applied the fitted trend on all the HCT spectra by dividing them with this calibration curve (assuming this calibration trend remains the same regardless of epoch). With this empirical correction we obtained the HCT spectral sequence shown in Fig.~\ref{fig:2004et_clear}, which we passed to the regular relative flux calibration routines described in Sect.~\ref{sec:Calib}.

\section{Model parameters}
In Table~\ref{tab:fit_pms} we show the physical parameters obtained for the various supernovae using the emulator.

\begin{table*}[]
    \centering
    \begin{tabular}{c c c c c c}
    Epoch [d] & $T_{\textrm{ph}}$ [K] & $v_{\textrm{ph}}$ [km/s] & $n$ & $\Theta$ [$10^8$ km / Mpc] & $\Theta / v_{\textrm{ph}}$ [d / Mpc] \\
    \hline
    \hline
    \multicolumn{6}{c}{\textbf{M61 -- NGC~4303}} \\
    \hline
    \hline
    \multicolumn{6}{c}{2008in} \\
    7.59 & 11955 & 8019 & 11.30 & 3.617 & 0.517 \\
    8.53 & 11081 & 7656 & 11.70 & 3.402 & 0.515 \\
    9.50 & 10542 & 7547 & 10.93 & 3.834 & 0.586 \\
    13.99 & 6948 & 6094 & 11.62 & 5.484 & 1.042 \\
    16.30 & 6570 & 6324 & 10.92 & 5.924 & 1.084 \\
    29.32 & 5962 & 3988 & 7.70 & 6.853 & 1.989 \\
    \hline
    \multicolumn{6}{c}{2020jfo} \\
    7.08 & 10665 & 10518.75 & 9.76 & 4.675 & 0.537 \\
    9.44 & 10176 & 8898.72 & 9.68 & 5.136 & 0.685 \\
    11.39 & 8984 & 8606.20 & 9.08 & 5.936 & 0.799 \\
    12.00 & 9552 & 8071.05 & 9.02 & 5.638 & 0.815 \\
    13.98 & 8215 & 8450 & 9.86 & 6.420 & 0.884 \\
    15.00 & 8156 & 7786 & 9.37 & 6.448 & 0.965 \\
    19.42 & 6563 & 6519 & 10.50 & 7.984 & 1.417 \\
    \hline
    \hline
    \multicolumn{6}{c}{\textbf{NGC~772}} \\
    \hline
    \hline
    \multicolumn{6}{c}{2003hl} \\
    15.28 & 10425 & 9049 & 10.15 & 3.610 & 0.462\\
    \hline
    \multicolumn{6}{c}{2003iq} \\
    8.98 & 10196 & 10557 & 11.58 & 3.429 & 0.376 \\
    16.19 & 8111 & 8935 & 10.38 & 4.561 & 0.591 \\
    20.79 & 6571 & 7629 & 11.01 & 5.530 & 0.839 \\
    29.04 & 5950 & 5904 & 8.76 & 6.048 & 1.186 \\
    \hline
    \hline
    \multicolumn{6}{c}{\textbf{NGC~922}} \\
    \hline
    \hline
    \multicolumn{6}{c}{2002gw} \\
    16.52 & 6861 & 7396 & 9.47 & 2.501 & 0.391\\
    20.12 & 6229 & 6464 & 8.21 & 2.723 & 0.484\\
    21.12 & 6221 & 6304 & 9.13 & 2.677 & 0.491\\
    \hline
    \multicolumn{6}{c}{2008ho} \\
    20.48 & 6242 & 5053 & 9.53 & 2.404 & 0.551\\
    25.43 & 5946 & 4423 & 8.00 & 2.607 & 0.682\\
    \hline
    \hline
    \multicolumn{6}{c}{\textbf{NGC~6946}} \\
    \hline
    \hline
    \multicolumn{6}{c}{2004et} \\
    10.44 & 10377 & 12401 & 11.06 & 19.818 & 1.866\\
    12.31 & 10308 & 11562 & 10.36 & 20.626 & 2.047\\
    14.30 & 10181 & 10604 & 9.68 & 21.557 & 2.334\\
    20.19 & 8505 & 8185 & 12.95 & 22.185 & 3.312\\
    24.00 & 7386 & 7708 & 13.25 & 26.994 & 4.064\\
    29.98 & 6294 & 6436 & 14.90 & 30.373 & 5.315\\
    \hline
    \multicolumn{6}{c}{2017eaw} \\
    11.78 & 9764 & 10388 & 14.69 & 16.971 & 1.891\\
    13.77 & 9450 & 10170 & 12.14 & 18.339 & 2.087\\
    15.74 & 8373 & 9975 & 11.18 & 20.771 & 2.410\\
    19.73 & 7199 & 8987 & 11.91 & 23.629 & 3.043\\
    21.73 & 6764 & 8569 & 11.88 & 25.082 & 3.388\\
    22.72 & 6550 & 8451 & 11.81 & 25.686 & 3.518\\
    24.72 & 6318 & 7866 & 11.46 & 26.551 & 3.907\\
    32.83 & 5946 & 6247 & 8.16 & 28.740 & 5.325\\
    \hline
    \end{tabular}
    \caption{Table of the inferred physical parameters for the various supernovae. $T_{\textrm{ph}}$ and $v_{\textrm{ph}}$ denote the photospheric temperature and velocity respectively, $n$ the density exponent, and $\Theta$ the estimated angular diameter of the supernovae.}
    \label{tab:fit_pms}
\end{table*}

\section{Light curve fits}
\label{Sec:lc_fits}
\begin{figure*}
    \centering
    \includegraphics[width=0.40\linewidth]{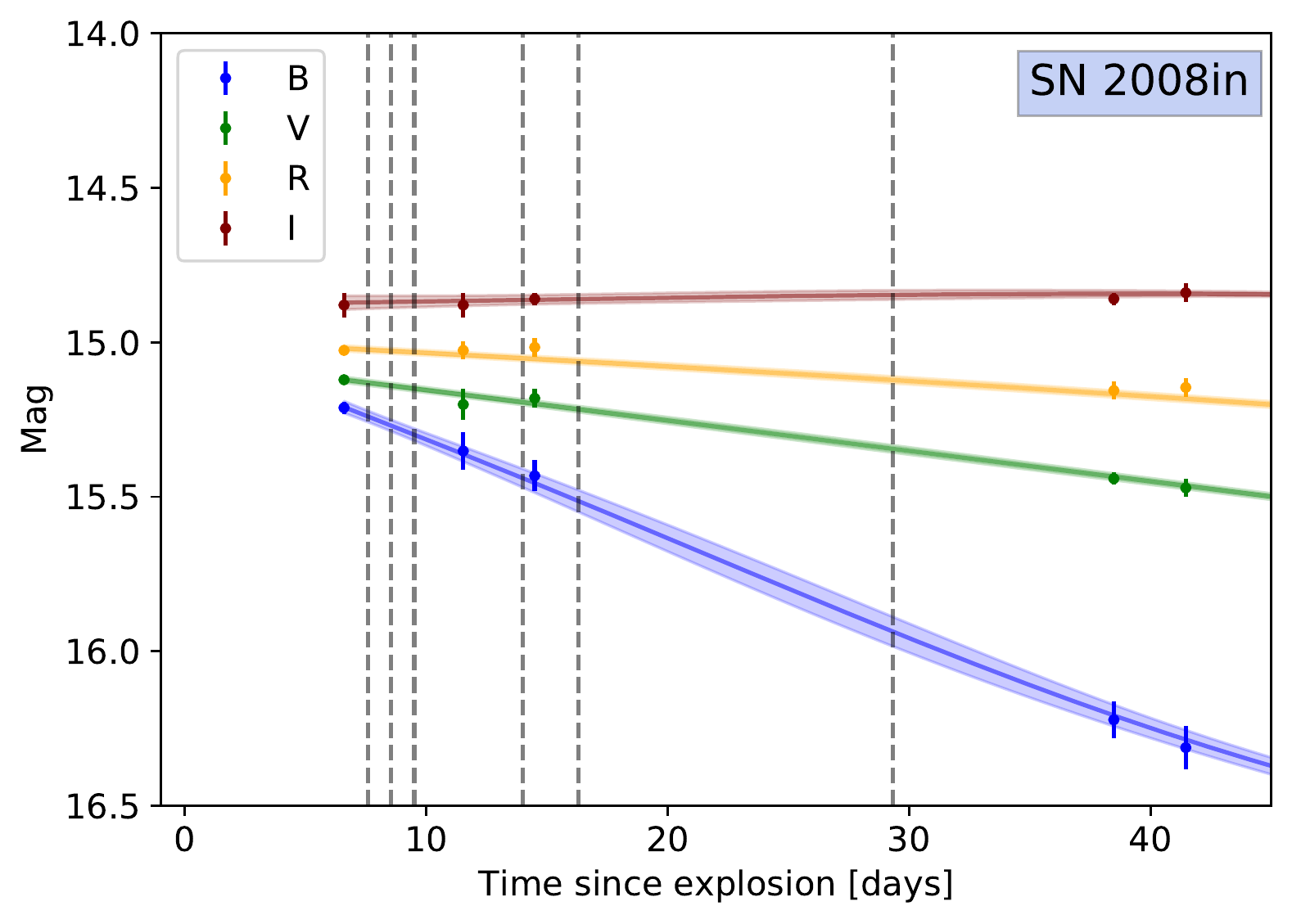}    
    \includegraphics[width=0.412\linewidth]{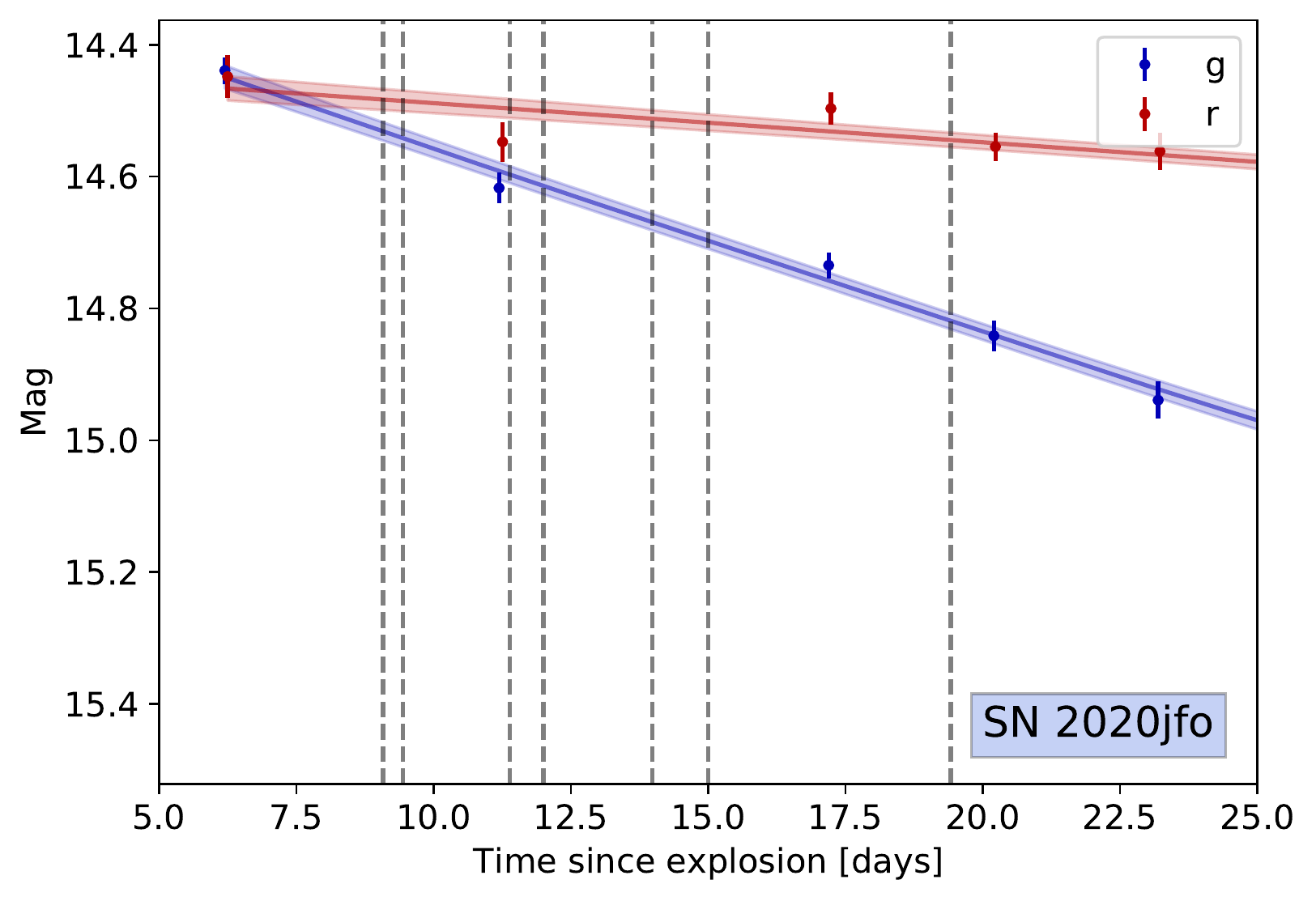}   
    \includegraphics[width=0.40\linewidth]{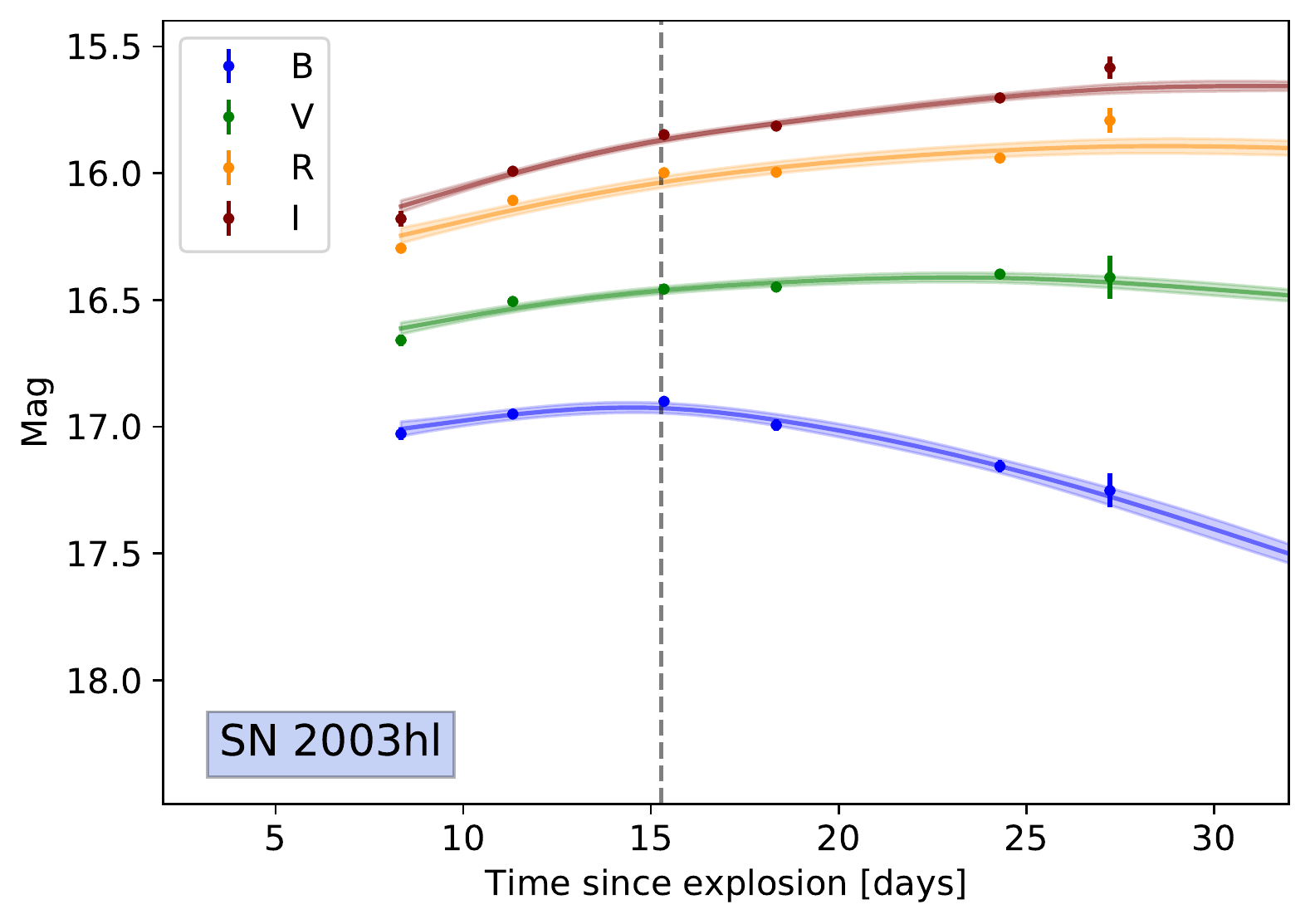}
    \includegraphics[width=0.407\linewidth]{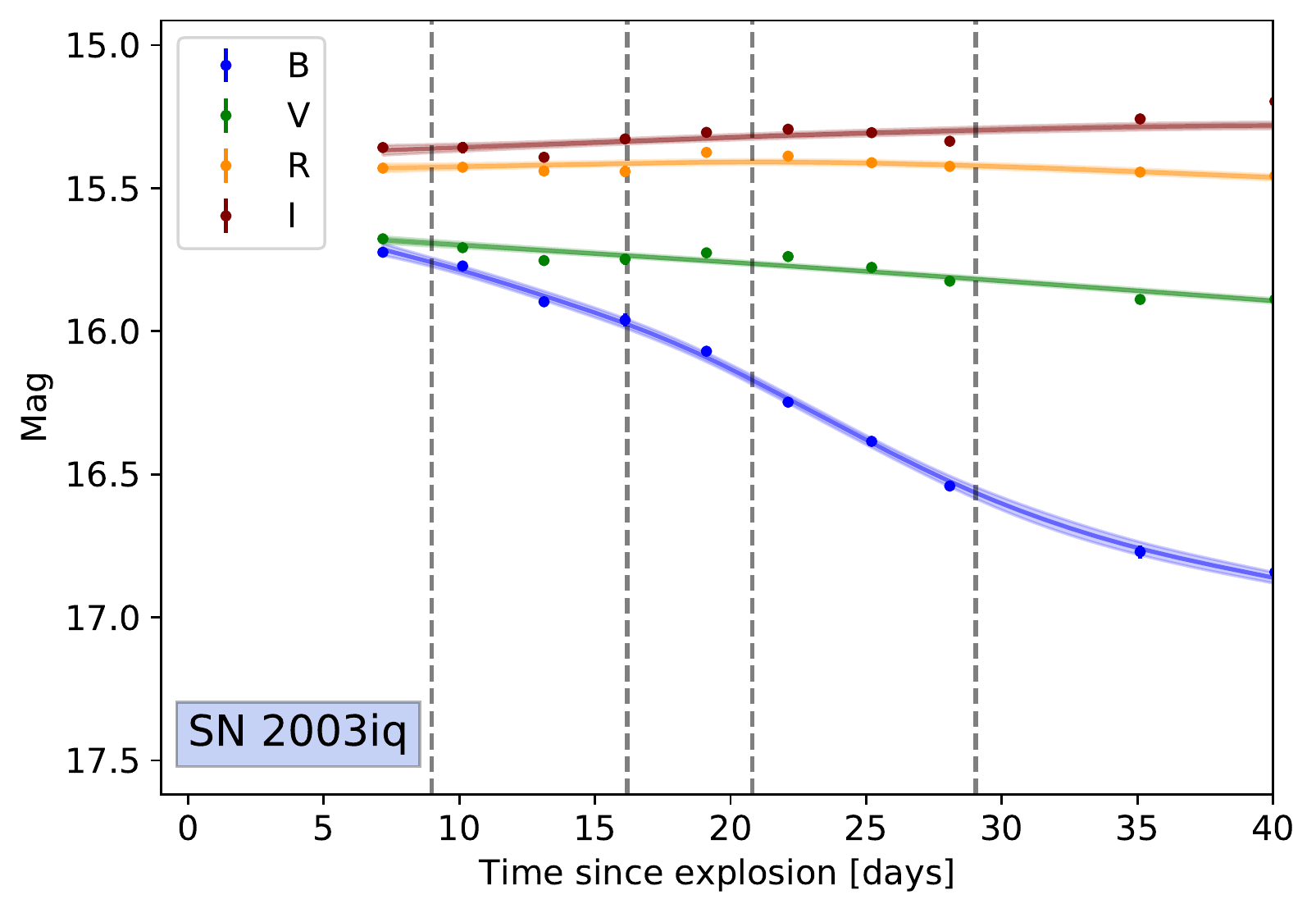}   
    \hspace*{-6pt}
    \includegraphics[width=0.42\linewidth]{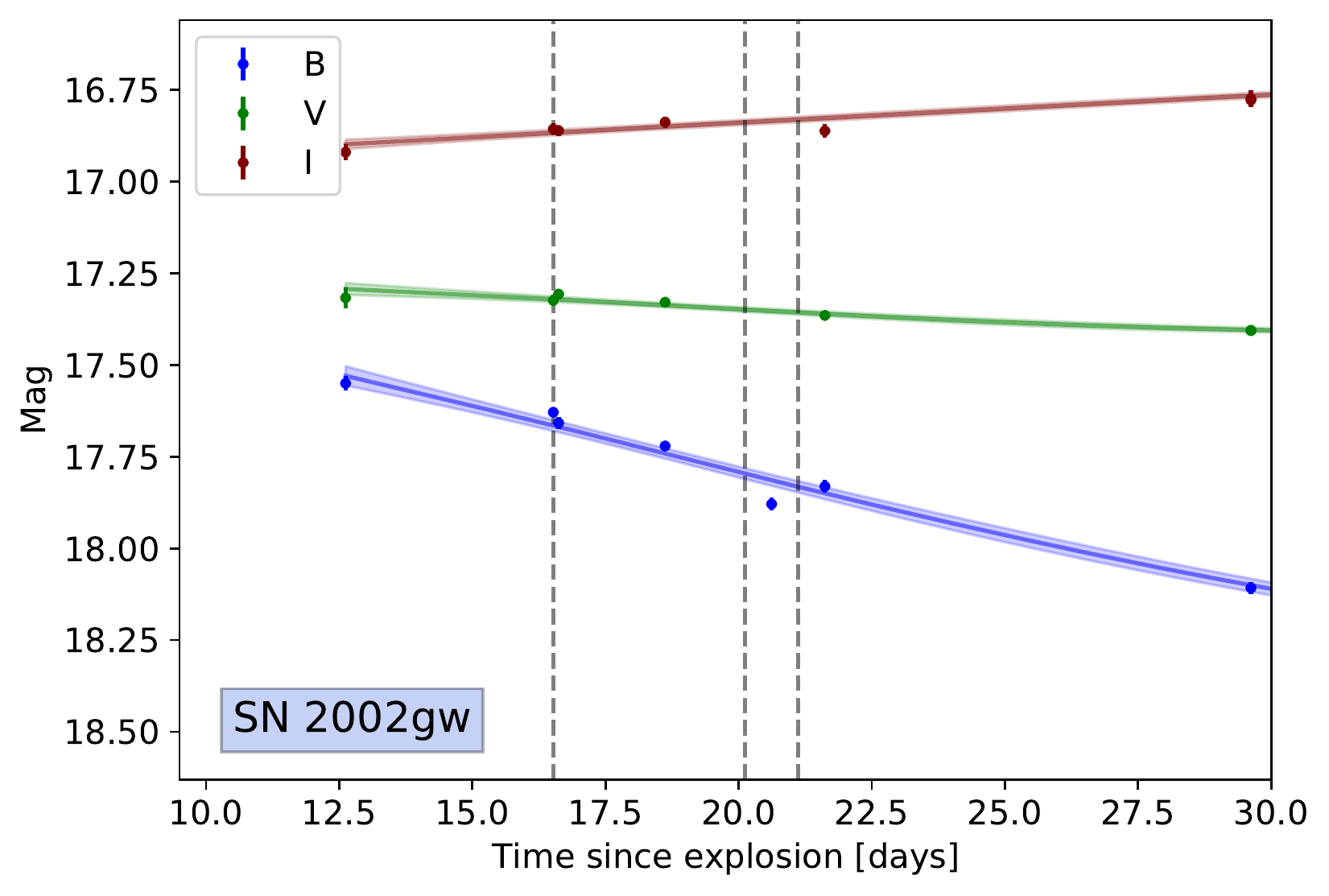}    
    \hspace*{-8pt}
    \includegraphics[width=0.406\linewidth]{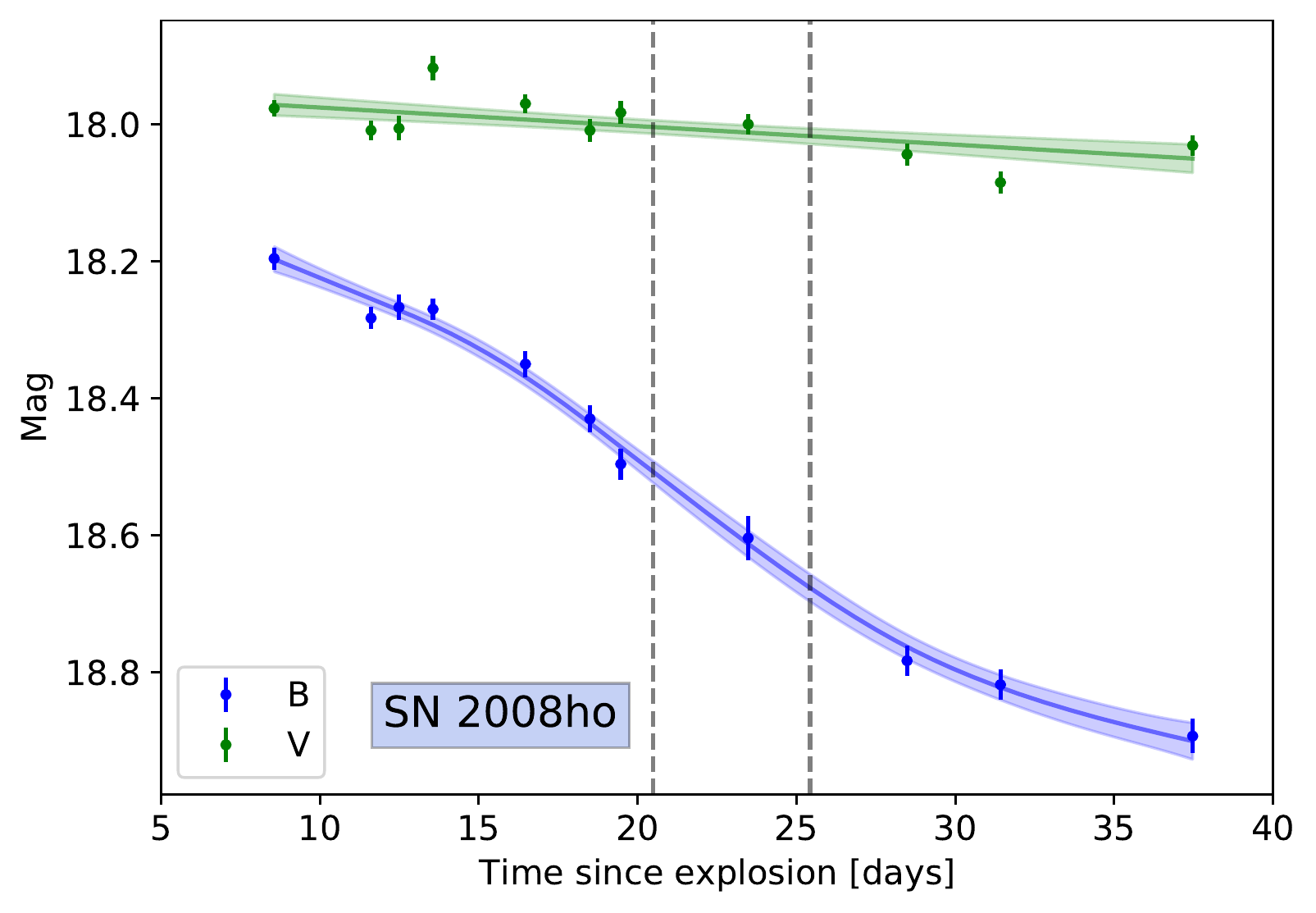}    
    \includegraphics[width=0.40\linewidth]{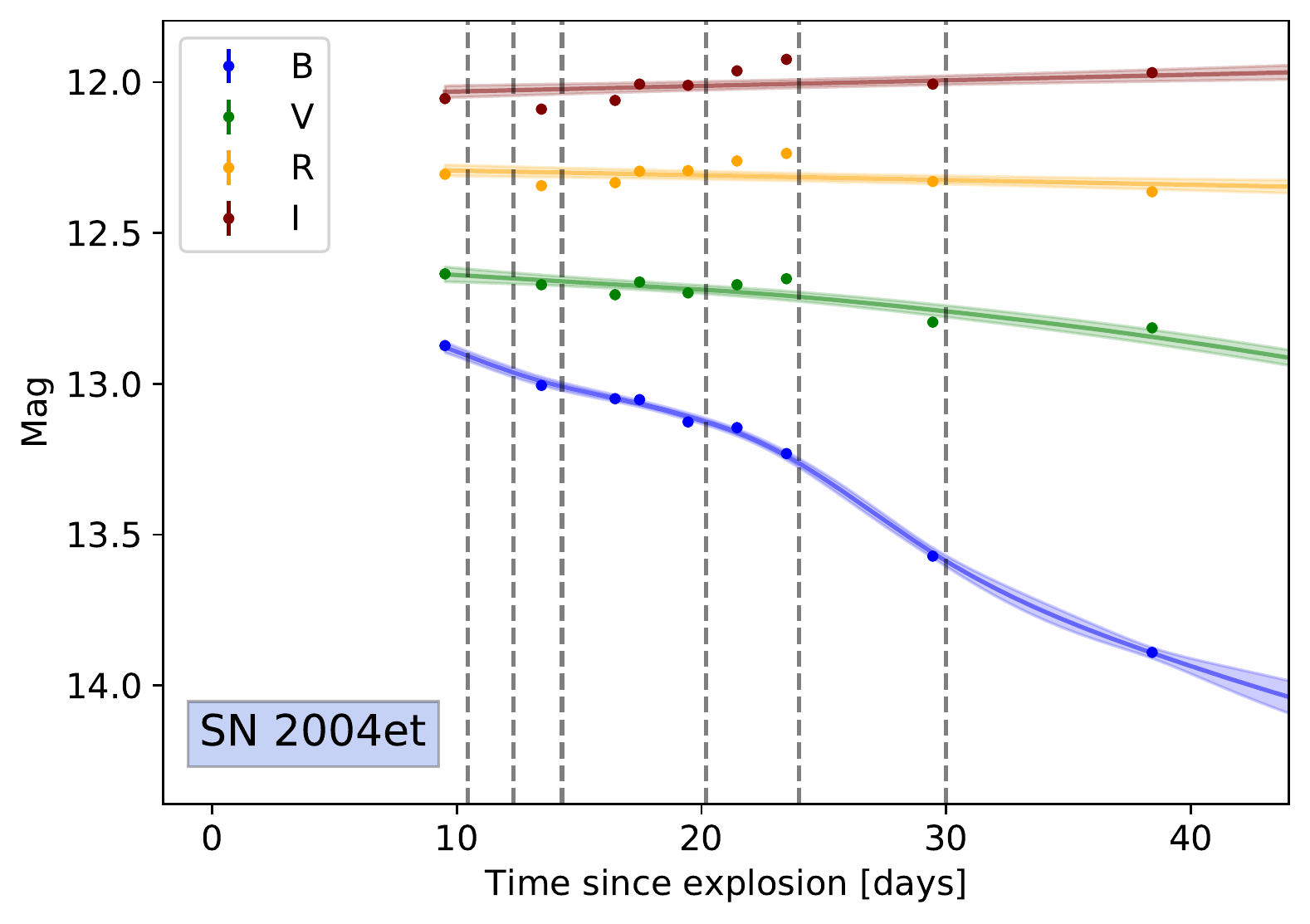}
    \includegraphics[width=0.40\linewidth]{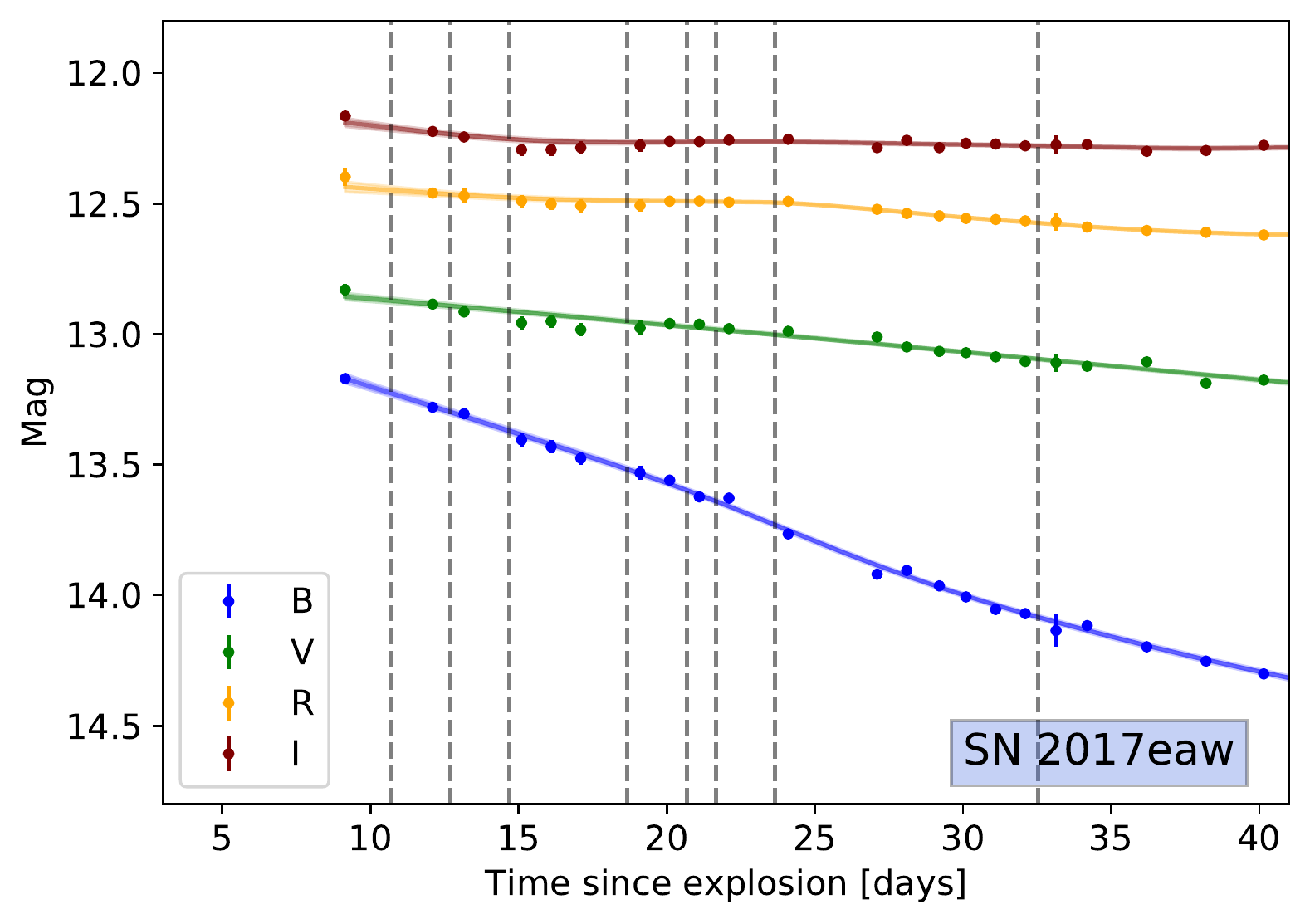}    
    \caption{The interpolated light curves of the individual supernovae. The vertical grey lines show the epochs of spectral observations.}
    \label{fig:lcs}
\end{figure*}

In Table~\ref{tab:magnitudes} we list the interpolated magnitudes required for the angular diameter calculations, while Fig.~\ref{fig:lcs} shows the individual light curve fits.

\begin{table*}[]
    \centering
    \begin{tabular}{c c c c c c c}
    Epoch [d] & $B$ & $V$ & $R$ & $I$ & $g$ & $r$ \\
    \hline
    \hline
    \multicolumn{7}{c}{\textbf{M61 -- NGC~4303}} \\
    \hline
    \hline
    \multicolumn{7}{c}{2008in} \\
    7.59 & 15.17(.02) & 15.11(.01) & 15.01(.01) & 14.87(.02) & \textemdash & \textemdash \\
    8.53 & 15.21(.02) & 15.12(.01) & 15.02(.01) & 14.87(.02) & \textemdash & \textemdash \\
    9.50 & 15.22(.02) & 15.12(.01) & 15.02(.01) & 14.87(.02) & \textemdash & \textemdash \\
    13.99 & 15.27(.02) & 15.14(.01) & 15.03(.01) & 14.87(.02) & \textemdash & \textemdash \\
    16.30 & 15.30(.02) & 15.15(.01) & 15.03(.01) & 14.87(.02) & \textemdash & \textemdash \\
    29.32 & 15.44(.02) & 15.19(.01) & 15.05(.01) & 14.86(.02) & \textemdash & \textemdash \\
    \hline
    \multicolumn{7}{c}{2020jfo} \\
    7.08 & \textemdash & \textemdash & \textemdash & \textemdash & 14.54(.02) & 14.48(.02) \\
    9.44 & \textemdash & \textemdash & \textemdash & \textemdash & 14.59(.01) & 14.50(.02) \\
    11.39 & \textemdash & \textemdash & \textemdash & \textemdash & 14.61(.01) & 14.50(.02) \\
    12.00 & \textemdash & \textemdash & \textemdash & \textemdash & 14.67(.01) & 14.51(.01) \\
    13.98 & \textemdash & \textemdash & \textemdash & \textemdash & 14.69(.01) & 14.52(.01) \\
    15.00 & \textemdash & \textemdash & \textemdash & \textemdash & 14.82(.01) & 14.54(.01) \\
    19.42 & \textemdash & \textemdash & \textemdash & \textemdash & 14.82(.01) & 14.54(.01) \\
    \hline
    \hline
    \multicolumn{7}{c}{\textbf{NGC~772}} \\
    \hline
    \hline
    \multicolumn{7}{c}{2003hl} \\
    15.28 & 16.93(.02) & 16.46(.01) & 16.04(.02) & 15.87(.01) & \textemdash & \textemdash \\
    \hline
    \multicolumn{7}{c}{2003iq} \\
    8.98 & 15.76(.02) & 15.75(.01) & 15.42(.01) & 15.33(.02) & \textemdash & \textemdash \\
    16.19 & 15.97(.02) & 15.72(.01) & 15.42(.01) & 15.33(.01) & \textemdash & \textemdash \\
    20.79 & 16.17(.02) & 15.74(.01) & 15.38(.01) & 15.30(.01) & \textemdash & \textemdash \\
    29.04 & 16.57(.02) & 15.82(.01) & 15.43(.01) & 15.33(.01) & \textemdash & \textemdash \\
    \hline
    \hline
    \multicolumn{7}{c}{\textbf{NGC~922}} \\
    \hline
    \hline
    \multicolumn{7}{c}{2002gw} \\
    16.52 & 17.66(.02) & 17.32(.01) & \textemdash & 16.87(.01) & \textemdash &  \textemdash\\
    20.12 & 17.80(.01) & 17.35(.01) & \textemdash & 16.84(.01) & \textemdash & \textemdash \\
    21.12 & 17.83(.02) & 17.36(.01) & \textemdash & 16.83(.01) & \textemdash & \textemdash \\
    \hline
    \multicolumn{7}{c}{2008ho} \\
    20.48 & 18.51(.01) & 18.00(.01) & \textemdash & \textemdash & \textemdash & \textemdash \\
    25.43 & 18.68(.01) & 18.02(.01) & \textemdash & \textemdash & \textemdash & \textemdash \\
    \hline
    \hline
    \multicolumn{7}{c}{\textbf{NGC~6946}} \\
    \hline
    \hline
    \multicolumn{7}{c}{2004et} \\
    10.44 & 12.91(.02) & 12.63(.01) & 12.29(.01) & 12.03(.02) & \textemdash & \textemdash \\
    12.31 & 12.96(.02) & 12.65(.01) & 12.29(.01) & 12.03(.02) & \textemdash & \textemdash \\
    14.30 & 13.00(.02) & 12.66(.01) & 12.30(.01) & 12.02(.01) & \textemdash & \textemdash \\
    20.19 & 13.13(.02) & 12.70(.01) & 12.31(.01) & 12.02(.01) & \textemdash & \textemdash \\
    24.00 & 13.27(.02) & 12.73(.01) & 12.32(.01) & 12.01(.01) & \textemdash & \textemdash \\
    29.98 & 13.58(.02) & 12.78(.01) & 12.34(.01) & 12.00(.01) & \textemdash & \textemdash \\
    \hline
    \multicolumn{7}{c}{2017eaw} \\
    11.78 & 13.24(.01) & 12.87(.01) & 12.45(.01) & 12.21(.01) & \textemdash & \textemdash \\
    13.77 & 13.30(.01) & 12.90(.01) & 12.46(.01) & 12.23(.01) & \textemdash & \textemdash \\
    15.74 & 13.37(.01) & 12.92(.01) & 12.48(.01) & 12.25(.01) & \textemdash & \textemdash \\
    19.73 & 13.51(.01) & 12.95(.01) & 12.49(.01) & 12.26(.01) & \textemdash & \textemdash \\
    21.73 & 13.60(.01) & 12.97(.01) & 12.49(.01) & 12.26(.01) & \textemdash & \textemdash \\
    22.72 & 13.64(.01) & 12.98(.01) & 12.49(.01) & 12.26(.01) & \textemdash & \textemdash \\
    24.72 & 13.73(.01) & 12.99(.01) & 12.50(.01) & 12.27(.01) & \textemdash & \textemdash \\
    32.83 & 14.08(.01) & 13.10(.01) & 12.57(.01) & 12.28(.01) & \textemdash & \textemdash \\
    \hline
    \end{tabular}
    \caption{Table of the interpolated magnitudes for the various supernovae. The values in the brackets denote the uncertainties of the estimates.}
    \label{tab:magnitudes}
\end{table*}

\end{appendix}

\end{document}